\documentclass[epj]{svjour}
\usepackage{graphics}
\usepackage{amsmath}
\usepackage{amsfonts}
\usepackage{amssymb}
\usepackage{float}
\usepackage{color}
\usepackage{graphicx}
\usepackage{graphics}
\usepackage{hyperref}
\usepackage{adjustbox,array}
\usepackage{enumitem} 
\usepackage{makecell}
\usepackage{mathrsfs}
\usepackage{multirow}
\newcommand{\gsim}{\gtrsim}
\newcommand{\lsim}{\lesssim}

\newcommand{\lf}{\left(}
\newcommand{\ri}{\right)}

\newcommand{\nn}{\nonumber}

\newcommand{\sqt}{\sqrt{2}}

\renewcommand{\lg}{\mathscr{L}} % Amplitude
\newcommand{\br}{\text{Br}}
\newcommand{\hc}{{\rm H.c.}}

\newcommand{\pb}{{\;{\rm pb}}}
\newcommand{\fb}{{\;{\rm fb}}}
\newcommand{\ab}{{\;{\rm ab}}}

\newcommand{\iab}{{\;{\rm ab}^{-1}}}

\newcommand{\gev}{{\;{\rm GeV}}}
\newcommand{\tev}{{\;{\rm TeV}}}

\newcommand{\beq}{\begin{equation}}
\newcommand{\eeq}{\end{equation}}
\newcommand{\bea}{\begin{eqnarray}}
\newcommand{\eea}{\end{eqnarray}}
\newcommand{\barr}{\begin{array}}
\newcommand{\earr}{\end{array}}
\newcommand{\bc}{\begin{center}}
\newcommand{\ec}{\end{center}}
\newcommand{\bit}{\begin{itemize}}
\newcommand{\eit}{\end{itemize}}
\newcommand{\ben}{\begin{enumerate}}
\newcommand{\een}{\end{enumerate}}
\newcommand{\al}{\alpha}
\newcommand{\bt}{\beta}

\newcommand{\sg}{\sigma}

\newcommand{\gm}{\gamma}

\newcommand{\lm}{\lambda}

\newcommand{\tauh}{{\tau_{\rm h}}}

\newcommand{\hsm}{{h_{\rm SM}}}
\newcommand{\ch}{H^\pm}

\newcommand{\wpm}{W^\pm}

\newcommand{\mh}{m_{h}}
\newcommand{\mch}{M_{H^\pm}}
\newcommand{\mhh}{M_{H}}
\newcommand{\ma}{M_{A}}
\newcommand{\mhha}{M_{H/A}}
\newcommand{\muf}{\mu_{\rm f}}
\newcommand{\mmuf}{\mu_{\rm f}^+ \mu_{\rm f}^- }

\newcommand{\tb}{\tan\beta}

\newcommand{\cb}{c_\beta}
\renewcommand{\sb}{s_\beta}

\newcommand{\mmu}      {{\mu^+ \mu^-}}
\newcommand{\nnu}      {\nu\bar{\nu}}
\newcommand{\ttt}      {\tilde{t}}
\newcommand{\bbb}      {\tilde{b}}

\newcommand{\elll}      {{\ell^+\ell^- }}
\newcommand{\met}      {{E_T^{\rm miss}}}

\newcommand{\nsg}{N_{\rm s}}
\newcommand{\nbg}{N_{\rm b}}
\newcommand{\dbg}{\delta_{\rm bg}}
\newcommand{\Dbg}{\Delta_{\rm B}}
\begin{document}

\title{Discovery Prospects for the Light Charged Higgs Boson Decay to an Off-Shell Top Quark and a Bottom Quark at Future High-Energy Colliders}
\author{Jinheung Kim\inst{1} \and Soojin Lee\inst{2} \and Prasenjit Sanyal\inst{2} \and  Jeonghyeon Song\inst{2} \and Daohan Wang\inst{3}% etc
}                     % Do not remove
\offprints{}          % Insert a name or remove this line
\institute{School of Physics, Korea Institute for Advanced Study, 85 Hoegi-ro, Dongdaemun-gu, Seoul 02455, Republic of Korea \and Department of Physics, Konkuk University, 120 Neungdong-ro, Gwangjin-gu, Seoul 05029, Republic of Korea \and Institute of High Energy Physics (HEPHY), Austrian Academy of Sciences (OeAW), Georg-Coch-Platz 2, A-1010 Vienna, Austria}
\date{Received: date / Revised version: date}
\abstract{
The charged Higgs boson ($H^\pm$) with a mass below the top quark mass remains a viable possibility within the Type-I two-Higgs-doublet model under current constraints. While previous LHC searches have primarily focused on the $H^\pm\to\tau^\pm\nu$ decay mode, the decay channel into an off-shell top quark and a bottom quark, $H^\pm \rightarrow t^*b$,  is leading or subleading for $H^\pm$ masses between 130 and 170 GeV. This study investigates the discovery potential of future colliders for this off-shell decay mode through pair-produced charged Higgs bosons decaying via $H^+H^-\rightarrow t^*b\tau\nu\rightarrow bbjj\tau\nu$. We perform signal-to-background analyses at the HL-LHC and a prospective 100 TeV proton-proton collider, employing cut-flow strategies and the Boosted Decision Tree method. However, due to the softness of the $b$ jets, signal significances fall below detection thresholds at these facilities. Extending our study to a multi-TeV muon collider (MuC), we demonstrate that a 3 TeV MuC achieves high signal significance, surpassing the $5\sigma$ threshold with an integrated luminosity of 1 ab$^{-1}$ and a 10\% background uncertainty. Specifically, for $M_{H^\pm} = 130$, 150, and 170 GeV, the significances are 13.7, 13.5, and 6.06, respectively. In contrast, a 10 TeV MuC requires 10 ab$^{-1}$ to achieve similar results. Our findings highlight the critical role of the MuC in probing the new signal channel $H^\pm\rightarrow t^*b$, offering a promising avenue for future charged Higgs boson searches involving off-shell top quarks.
\PACS{
      {14.80.Fd}{Charged Higgs bosons}   \and
      {14.65.Ha}{top quarks}
     } % end of PACS codes
} %end of abstract
\maketitle
\section{Introduction}
\label{intro}

The milestone discovery of the Higgs boson at the LHC~\cite{ATLAS:2012yve,CMS:2012qbp} 
seemingly completes the Standard Model (SM), 
yet the quest for a new particle physics theory beyond the SM (BSM) continues. 
This pursuit is driven by unresolved fundamental questions of the Universe, 
such as the naturalness problem, fermion mass hierarchy, baryogenesis, non-zero neutrino masses, 
and the identity of dark matter. 
High-energy collider experiments are indispensable in this quest, 
offering the ability to directly study fundamental particles in a highly controlled environment 
and providing complementary insights to cosmological and dark matter searches.

One of the most promising BSM signals 
at high-energy colliders 
is the presence of a light charged Higgs boson with a mass below the top quark mass $m_t$. 
This possibility remains viable under current constraints within the Type-I and Type-X\footnote{In Type-II and Type-Y 2HDM, the 
charged Higgs boson  is tightly constrained to be as heavy 
as $\mch\gsim 800\gev$ due to the measurements of the inclusive $B$-meson 
decay into $X_s \gamma$~\cite{Misiak:2020vlo}.} 
two-Higgs-doublet model (2HDM)~\cite{Aoki:2009ha,Branco:2011iw,Craig:2013hca,Wang:2022yhm,Shen:2022yuo,Kanemura:2022ldq,Lee:2022gyf}, 
three-Higgs doublet model~\cite{Akeroyd:2018axd}, next-to-2HDM~\cite{Abouabid:2021yvw}, 
lepton-specific Inert doublet model~\cite{Han:2021gfu}, and scalar-triplet model~\cite{Ferreira:2021bdj}.

A well-studied prototype of these BSM scenarios is the 2HDM, which extends the SM Higgs sector by introducing two complex $SU(2)_L$ Higgs doublets, $\Phi_1$ and $\Phi_2$, both carrying hypercharge $Y=+1$. These fields acquire vacuum expectation values, $v_1$ and $v_2$, whose ratio, $\tan\beta = v_2/v_1$, is a key parameter shaping the model’s phenomenology. The physical Higgs spectrum consists of five states: the lighter \textit{CP}-even scalar $h$, the heavier \textit{CP}-even scalar $H$, the \textit{CP}-odd pseudoscalar $A$, and a pair of charged Higgs bosons, $H^\pm$. The existence of a light $H^\pm$ offers rich collider phenomenology, motivating dedicated experimental searches.  

Experimental efforts to discover a light charged Higgs boson span various colliders, including the LHC and future lepton colliders. 
For the decay $H^\pm \rightarrow \tau^\pm \nu$, 
various production channels have been explored, 
such as $t \rightarrow H^+ b$~\cite{Abbaspour:2018ysj,Demir:2018iqo,ATLAS:2018gfm,Sanyal:2019xcp,Sirunyan:2019hkq,Ghosh:2022wbe},
$pp \to H^\pm \varphi^0$~\cite{Kim:2023lxc}, $pp \rightarrow H^\pm A$~\cite{Kanemura:2011kx},
$pp \rightarrow H^+ H^-$~\cite{Duarte:2024zeh},
$cs/cb \rightarrow \ch$~\cite{Hernandez-Sanchez:2012vxa,Hernandez-Sanchez:2020vax},
$pp \rightarrow W^{\pm *}W^{\pm *} \rightarrow H^\pm H^\pm$~\cite{Aiko:2019mww}, and
$e^+ e^- \to H^+ H^-$~\cite{Coniavitis:2007me}.
Here, $ \varphi^0$ denotes a new \textit{CP}-even neutral Higgs boson.
For the $H^\pm\rightarrow c b/cs$ mode, the production channels of $t \rightarrow H^+  b$~\cite{ATLAS:2013uxj,CMS:2015yvc,Akeroyd:2018axd,CMS:2018dzl,ATLAS:2021zyv,CMS:2020osd,Akeroyd:2022ouy} and $e^+ e^-\to H^+ H^-$~\cite{Hou:2021qff} have been considered.
The decays $H^\pm\rightarrow W^\pm \varphi^0/W^\pm A$ have been extensively studied
for production channels such as 
$t \rightarrow H^\pm b$~\cite{Dermisek:2012cn,Arhrib:2020tqk,Hu:2022gwd,Fu:2023sng}, 
$pp \rightarrow H^\pm \varphi^0$~\cite{Arhrib:2017wmo,Mondal:2021bxa,Arhrib:2022inj,Kim:2022hvh,Kim:2022nmm,Bhatia:2022ugu,Li:2023btx,Mondal:2023wib,Cheung:2022ndq}, 
$pp \rightarrow H^\pm A$~\cite{Arhrib:2021xmc,Arhrib:2021yqf}, 
$e^+ e^- \rightarrow H^+ H^-$~\cite{Kausar:2020ims,Barik:2022dvi,Ouazghour:2023plc},
$pp \rightarrow H^+ H^-$~\cite{Arhrib:2021xmc,Arhrib:2021yqf,Shen:2022yuo}, 
$pp \rightarrow W^{\pm *}W^{\pm *} \rightarrow H^\pm H^\pm$~\cite{Arhrib:2019ywg}, 
$pp \rightarrow H^\pm W^\mp$~\cite{Krab:2022lih}, $pp \rightarrow H^\pm h h $~\cite{Kang:2022mdy}, and the associated productions of $\elll \to \ch \tau^\mp\nu , tb\ch, \wpm H^\mp \varphi$~\cite{Ouazghour:2024twx}.

Despite these extensive studies, the decay of a light charged Higgs boson into an off-shell top quark (denoted as $t^*$) and a bottom quark remains largely unexplored in terms of its discovery potential at future colliders. This decay mode is particularly significant in the Type-I 2HDM, where all Yukawa couplings of $H^\pm$ share a common $1/\tan\beta$ dependence. Consequently, the branching ratio for $H^\pm \rightarrow f \bar{f}'$ is dictated primarily by fermion masses rather than $\tan\beta$ itself. If kinematically accessible, decays to heavier fermions dominate. In particular, $H^\pm \rightarrow t^* b$ emerges as the leading decay mode for $ M_{H^\pm} \in [135\gev, m_t]$, with $H^\pm \rightarrow \tau^\pm \nu$ as the second most prominent channel. Even for $\mch$ slightly below 135 GeV, this mode remains among the most significant, highlighting its importance over the mass range $\mch \in [130,170] \gev$. This motivates a detailed study of $H^\pm \to t^*b$ and its potential as a discovery channel at future colliders.

Previous research on $H^\pm \rightarrow t^* b$ has primarily focused on the branching ratios of the related $H^\pm \to b \bar{b} W^\pm$ decay, mainly within the Minimal Supersymmetric Standard Model, i.e., Type-II of the 2HDM~\cite{Moretti:1994ds,Djouadi:1995gv,Borzumati:1998xr,Bi:1999hp,Djouadi:2005gj}. Although Ref.~\cite{Ma:1997up} demonstrated that this decay mode could be enhanced for $M_{H^\pm} > 140 \gev$ and $\tan \beta \lesssim 1$, their analysis was restricted to parton-level and focused only on the Tevatron.

To advance charged Higgs boson searches, it is essential to thoroughly investigate the observability of the $H^\pm \to t^* b$ decay mode at future high-energy colliders. This need forms the primary motivation for our work. We provide the first comprehensive analysis of the discovery potential for the $H^\pm \to t^* b$ decay mode across multiple future collider scenarios.

For the production of $H^\pm$ at high-energy colliders, we focus on pair production. This process depends solely on $M_{H^\pm}$, making it a robust and unambiguous choice for our study. Traditional searches use the channel $pp \to t \bar{t}$, followed by $t \to b H^+$ and $H^\pm \to \tau^\pm \nu$~\cite{Abbaspour:2018ysj,Demir:2018iqo,ATLAS:2018gfm,Sanyal:2019xcp,Sirunyan:2019hkq,Ghosh:2022wbe}, but this approach becomes highly inefficient in the large $\tan\beta$ limit.
Current bounds already favor $\tan\beta$ values above about 6, which suppress $\br(t \to b H^+)$ to below approximately $2 \times 10^{-3}$ for $M_{H^\pm} = 150 \gev$. If $\tan\beta = 10$, this branching ratio further reduces to around $7 \times 10^{-4}$. These values were obtained from our calculation using {\small\sc 2HDMC}~\cite{Eriksson:2009ws}.
 Other production channels involving decays of additional BSM Higgs bosons are sensitive to their masses and $\tan\beta$. In contrast, pair production avoids these model-dependent complications, providing a clearer path for investigation.

Motivated by these considerations, we will rigorously investigate the discovery potential of the HL-LHC and a 100 TeV $pp$ collider for the signal $H^+H^-\rightarrow t^*b\tau\nu\rightarrow bbjj\tau\nu$, employing both cut-flow strategies and the Boosted Decision Tree (BDT) method. However, the analysis reveals that these efforts are ultimately unsuccessful due to the softness of the $b$ jets. Given these challenges, our research shifts focus to the multi-TeV muon collider (MuC). The MuC is a powerful tool for BSM searches~\cite{Capdevilla:2020qel,Bandyopadhyay:2021pld,Sen:2021fha,Asadi:2021gah,Huang:2021nkl,Choi:1999kn,Han:2020uak,Han:2021udl,Jueid:2021avn,Han:2022edd,Black:2022qlg,Belfkir:2023vpo,Jueid:2023zxx}, offering a clean collision environment, higher energy reach, reduced beamstrahlung, and efficient energy usage. The prospects of the MuC program have been significantly enhanced by recent advancements in addressing critical challenges, such as cooling muon beams~\cite{Antonelli:2013mmk,Antonelli:2015nla} and reducing beam-induced backgrounds (BIBs)~\cite{Collamati:2021sbv,Ally:2022rgk}. The multi-TeV collision energy at the MuC promises a higher boost for the $b$ jets, potentially enhancing their detectability and opening new avenues for exploration.

In our study, we conduct a signal-to-background analysis for two MuC configurations: $\sqrt{s}=3\text{ TeV}$ with a total integrated luminosity of $1\text{ ab}^{-1}$ and $\sqrt{s}=10\text{ TeV}$ with a total integrated luminosity of $10\text{ ab}^{-1}$. This analysis aims to demonstrate that the full mass range of $M_{H^\pm} \in [130,170]\text{ GeV}$ can achieve high signal significance, surpassing the $5\sigma$ discovery threshold. These findings represent our main contributions to the study of the light charged Higgs boson.

The paper is organized as follows. 
In Sec.~\ref{sec-review}, we briefly review the Type-I 2HDM with \textit{CP} invariance and softly broken $Z_2$ parity. Based on the results of random scans incorporating theoretical and experimental constraints, we investigate the characteristic features of the allowed parameters and suggest the golden channel to probe the unexplored $H^\pm \rightarrow t^* b$ decay mode, $H^+H^-\rightarrow t^*b\tau\nu\rightarrow bbjj\tau\nu$. Section \ref{sec-LHC} deals with the signal-to-background analysis at the HL-LHC and a 100 TeV $pp$ collider, incorporating comprehensive cut-based analysis and the BDT. In Sec.~\ref{sec-MuC}, we turn to the multi-TeV MuC and perform the signal-to-background analysis. Conclusions are presented in Sec.~\ref{sec-conclusions}.

\section{Brief review of the light charged Higgs boson in Type-I 2HDM}
\label{sec-review}

The 2HDM introduces two complex $SU(2)_L$ Higgs doublet scalar fields, $\Phi_1$ and $\Phi_2$, with hypercharge $Y=+1$~\cite{Branco:2011iw}:
\bea
\label{eq-phi:fields}
\Phi_i = \left( \begin{array}{c} w_i^+ \\[3pt]
\dfrac{v_i + h_i + i \eta_i }{ \sqrt{2}}
\end{array} \right), \quad i=1,2,
\eea
where $v_1$ and $v_2$ denote the non-zero vacuum expectation values of $\Phi_1$ and $\Phi_2$, respectively. 
The ratio of $v_2$ to $v_1$ defines the mixing angle $\beta$ through
$\tan\beta =v_2/v_1$. 
The electroweak symmetry is spontaneously broken by $v =\sqrt{v_1^2+v_2^2} \approx246\gev $.

To prevent flavor-changing neutral currents at the tree level, 
a discrete $Z_2$ symmetry is imposed, under which $\Phi_1 \rightarrow \Phi_1$ and $\Phi_2 \rightarrow -\Phi_2$~\cite{Glashow:1976nt,Paschos:1976ay}. 
Assuming \textit{CP} invariance and allowing for softly broken $Z_2$ parity, the scalar potential is defined as follows:
\bea
\label{eq-VH}
V = && m^2 _{11} \Phi^\dagger _1 \Phi_1 + m^2 _{22} \Phi^\dagger _2 \Phi_2
-m^2 _{12} ( \Phi^\dagger _1 \Phi_2 + \hc) \\ \nn
&& + \frac{1}{2}\lambda_1 (\Phi^\dagger _1 \Phi_1)^2
+ \frac{1}{2}\lambda_2 (\Phi^\dagger _2 \Phi_2 )^2
\\ \nn
&& 
+ \lambda_3 (\Phi^\dagger _1 \Phi_1) (\Phi^\dagger _2 \Phi_2)
+ \lambda_4 (\Phi^\dagger_1 \Phi_2 ) (\Phi^\dagger _2 \Phi_1) \\ \nn
&& + \frac{1}{2} \lambda_5
\left[
(\Phi^\dagger _1 \Phi_2 )^2 +  \hc
\right].
\eea

In 2HDM, there are five physical Higgs bosons: 
the lighter \textit{CP}-even scalar $h$, the heavier \textit{CP}-even scalar $H$, the \textit{CP}-odd pseudoscalar $A$, 
and a pair of charged Higgs bosons $H^\pm$. 
These physical states are related to the weak eigenstates through the following transformations~\cite{Song:2019aav}:
\bea
\left(
\begin{array}{c}
h_1 \\ h_2
\end{array}
\right) &=&
\mathbb{R}(\al)
\left(
\begin{array}{c}
H \\ h
\end{array}
\right),
\quad
\left(
\begin{array}{c}
\eta_1 \\ \eta_2
\end{array}
\right) =
\mathbb{R}(\beta)
\left(
\begin{array}{c}
z^0 \\ A
\end{array}
\right)
, \\ \nn
\left(
\begin{array}{c}
w_1^\pm \\ w_2^\pm
\end{array}
\right) &=&
\mathbb{R}(\bt)
\left(
\begin{array}{c}
w^\pm \\ H^\pm
\end{array}
\right),
\eea
where \( z^0 \) and \( w^\pm \) are the Goldstone bosons that are absorbed to become the longitudinal components of the \( Z \) and \( W \) bosons, respectively. 
The rotation matrix $\mathbb{R}(\theta)$ is  defined as
\bea
\mathbb{R}(\theta) = \left(
\begin{array}{cr}
\cos \theta & -\sin\theta \\ \sin\theta & \cos\theta
\end{array}
\right).
\eea
The SM Higgs boson $\hsm$ is a linear combination of $h$ and $H$, specifically as $\hsm = \sin (\beta-\alpha) h + \cos(\beta-\alpha) H$. 
In this study, we focus on the normal scenario where $h$ is the observed Higgs boson with a mass of 125 GeV.

According to the assignment of $Z_2$ parity to the right-handed fermions, 
the model has four variants,  Type-I, Type-II, Type-X, and Type-Y. 
The mass of $\ch$ in Type-II and Type-Y is heavily constrained 
by the measurements of the inclusive $B$-meson decay into $X_s\gamma$, 
which requires $\mch \gsim 800\gev$~\cite{Misiak:2020vlo}. 
Only Type-I and Type-X allow for the existence of charged Higgs bosons lighter than the top quark.
Since we aim to study the $\ch\to t^* b$ decay, we focus on the Type-I model.

The model has six physical parameters of $\mch$, $\mhh$,  $\ma$, $m_{12}^2$, $\tb$, and $\sin (\beta-\al)$.
To focus on the key features, we employ two popular conditions: the Higgs alignment limit for the SM-like Higgs boson~\cite{Carena:2013ooa,Celis:2013rcs,Cheung:2013rva,Bernon:2015qea,Chang:2015goa,Das:2015mwa,Kanemura:2021dez} and the mass degeneracy of $H$ and $A$ for the electroweak precision data~\cite{Kanemura:2011sj,Chang:2015goa,Chen:2019pkq}. 
The Higgs alignment limit precludes the decay channel $H^\pm \to W^\pm h$.
Furthermore, we restrict our analysis to the scenario where the charged Higgs boson is lighter than $H$ and $A$.\footnote{If $\mhha < \mch$, the decay modes $\ch \rightarrow W^{\pm (*)}H/A$ become accessible~\cite{Sanyal:2023pfs}.}
In this scenario, the charged Higgs boson decays exclusively into fermion pairs, 
ensuring that the $\ch \rightarrow t^* b$ mode maintains a significant branching ratio. 

In summary, our model configuration is as follows:
\bea
\label{eq-focus:paper}
\text{Type-I:} && \mh=125\gev,\quad \sin (\beta-\alpha)=1, \\ \nn  &&\mch < \mhha (=\mhh=\ma) .
\eea
In this setup, the quartic couplings in \autoref{eq-VH} are~\cite{Das:2015mwa}:
\bea
\label{eq:quartic}
\lm_1 &=& \frac{1}{v^2 }
\left[
\lf \tan\beta \ri^2 (\mhh^2 - M^2) -  m_h^2 
\right], \\ \nn
\lm_2 &=& \frac{1}{v^2 }
\left[ \mh^2 + \frac{1}{\tb^2} \lf \mhh^2-M^2\ri
\right], \\ \nn
\lm_3 &=& 
\frac{1}{v^2}
\left[ \mh^2 +2 \mch^2 -\mhh^2-M^2
\right],
 \\[3pt] \nn
\lm_4 &=& 
\frac{1}{v^2}
\left[
\ma^2- 2 \mch^2 + M^2
\right], \\[3pt] \nn
\lm_5 &=&
\frac{1}{v^2}
\left[
M^2-\ma^2
\right], 
\eea
where $M^2 = m_{12}^2/(\sb\cb)$.

The Yukawa couplings of the charged Higgs boson to the SM fermions in Type-I 2HDM are given by
\begin{align}
\lg_{\rm Yuk} &=
-  \frac{1}{\tan\beta}
\left\{
\dfrac{\sqrt2V_{ud}}{v } H^+  \overline{u}
\left(m_u  {P}_- -  m_d  {P}_+\right)d 
\right.
\\ \nn & \left.
-\dfrac{\sqt m_\tau}{v}H^+ \overline{\nu}_L\tau_R^{}
+\hc
\right\},
\end{align}
where $P_\pm =(1\pm\gm^5)/2$.
Given that these Yukawa couplings share a common factor of $1/\tan\beta$,
the branching ratio of $\ch \rightarrow f\bar{f}$ depends solely on the fermion mass,
independent of $\tan\beta$.
As a result, when kinematically allowed, the decay of $H^\pm \to t^{(*)}b$ becomes overwhelmingly dominant.

We also present the gauge interactions of a pair of charged Higgs bosons,
crucial for the pair production at high-energy colliders~\cite{Branco:2011iw}:
\begin{align}
\label{eq-Lg:gauge-H+H-}
\lg_{\rm gauge} =
& ~i\left[
e A_\mu + \frac{g (s_W^2-c_W^2)}{2 c_W} Z_\mu 
\right]
\left( H^+ \stackrel{\leftrightarrow}{\partial^\mu} H^- \right)
\\ \nn 
& + H^+ H^-\left[
\frac{g^2}{2}W^{-\mu}W^+_\mu +  e^2 A^\mu A_\mu  \right.
\\ \nn & \left.+ 
\frac{g^2 (s_W^2-c_W^2)^2}{4 c_W^2} Z^\mu Z_\mu 
+ \frac{e g}{c_W (s_W^2-c_W^2)}A^\mu Z_\mu 
\right]  ,
\end{align} 
where $s_W =\sin\theta_W$, $c_W =\cos\theta_W$, $\theta_W$ is the electroweak mixing angle,
and $f\stackrel{\leftrightarrow}{\partial^\mu} g = f \partial^\mu g - g \partial^\mu f$.

To study the characteristics of the permissible parameter space for the light charged Higgs boson
that predominantly decays into $t^* b$, we consider three cases of $\mch=130,~150,~170\gev$ 
and perform a random scan within the following parameter ranges:
\begin{align}
\mhha &\in [150,1000]\gev, \quad m_{12}^2 \in [0, 10^5] \gev^2, \\ \nn  \tb &\in [1,50].
\end{align}
The scan is conducted while imposing both theoretical requirements and experimental constraints.

For the theoretical requirements, we enforce conditions ensuring vacuum stability~\cite{Ivanov:2008cxa,Barroso:2012mj,Barroso:2013awa}, a bounded-from-below Higgs potential~\cite{Ivanov:2006yq}, tree-level unitarity in scalar-scalar scatterings~\cite{Branco:2011iw,Arhrib:2000is}, and perturbativity of the Higgs quartic couplings~\cite{Chang:2015goa}. 
Furthermore, we require the parameter points to satisfy the current best-fit values for the Peskin-Takeuchi electroweak oblique parameters~\cite{Peskin:1991sw} in the 2HDM framework~\cite{He:2001tp,Grimus:2008nb}: $S=-0.04 \pm 0.10$, $T= 0.01 \pm 0.12$, and $U=-0.01\pm 0.09$~\cite{ParticleDataGroup:2024cfk}. The correlations among these oblique parameters are properly incorporated in our analysis.
These conditions are evaluated using the public code \textsc{2HDMC}~\cite{Eriksson:2009ws}. Additionally, we require that the cutoff scale exceed $10\tev$, where the cutoff scale is defined as the energy level at which any of the conditions for tree-level unitarity, perturbativity, or vacuum stability is violated~\cite{Kim:2023lxc}. The evolution of the model parameters via the renormalization group equations is facilitated using the public code \textsc{2HDME}~\cite{Oredsson:2018yho,Oredsson:2018vio}.

Regarding experimental constraints, we incorporated measurements at the 95\% confidence level, encompassing the Higgs precision data obtained at the LHC, the inclusive $B$-meson decay to $X_s \gamma$~\cite{Mahmoudi:2009zx,Arbey:2017gmh,Sanyal:2019xcp,Misiak:2017bgg}, 
$B \to K^* \gamma$~\cite{Belle:2017hum}, $B_s \to \phi \gamma$~\cite{Belle:2014sac},
and direct search bounds from LEP, Tevatron, and LHC experiments. 
For the Higgs precision and direct search constraints, we employed the public code \textsc{HiggsTools}-v1.2~\cite{Bahl:2022igd}.

\begin{figure}
\resizebox{0.49\textwidth}{!}{%
  \includegraphics{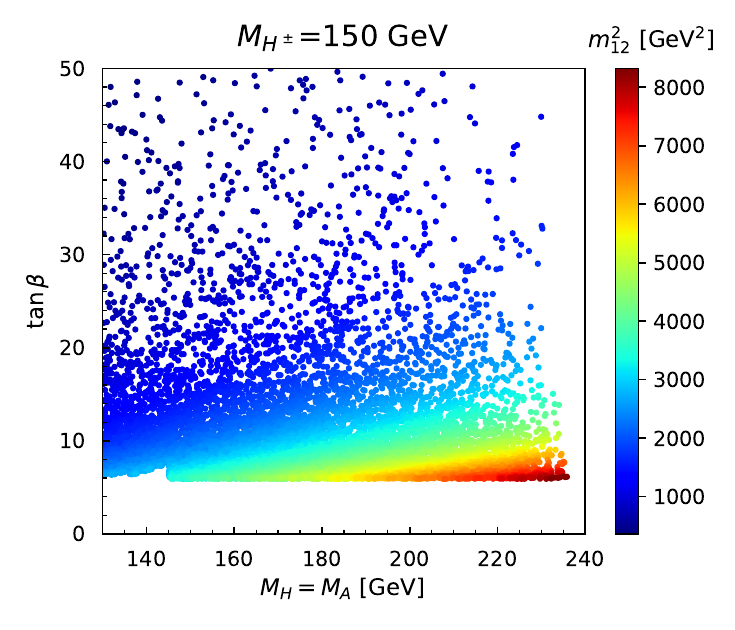}
}
%\vspace{5cm}       % Give the correct figure height in cm
\caption{Allowed parameter space of $(\mhha,\tan\beta)$ for $\mch=150\gev$.
The color code denotes $m_{12}^2$.}
\label{fig-mHc150-mHmA-tb}      % Give a unique label
\end{figure}

In \autoref{fig-mHc150-mHmA-tb}, we present the allowed parameter points in the $(\mhha, \tan\beta)$ plane for a fixed $\mch = 150\gev$. The color code represents $m_{12}^2$.  
A key feature of the results is the upper bound on the masses of $H$ and $A$, with $\mhha \lsim 236\gev$ for $\mch=150\gev$. 
This is a general characteristic of scenarios with a light charged Higgs boson, primarily driven by the stringent theoretical constraints from perturbativity and unitarity. Specifically, for a light $H^\pm$, the quartic couplings $\lambda_3$ and $\lambda_4$ increase significantly with $\mhha$ (see \autoref{eq:quartic}). As a result, excessively heavy $\mhha$ values would violate perturbativity and unitarity constraints.
Although our main focus in this paper is the light charged Higgs boson, the relatively low upper bounds on $\mhha$ indicate promising discovery prospects for the heavy neutral Higgs bosons at high-energy colliders. Additionally, we observe a lower bound on $\tan\beta$, specifically $\tan\beta \gsim 6$.

Now, let us identify the optimal production mechanism at the LHC for the light charged Higgs boson.  
LHC searches have primarily focused on its production through top quark decay, 
$t \rightarrow b H^+$, followed by $H^\pm \rightarrow \tau^\pm\nu$~\cite{ATLAS:2018gfm,Sirunyan:2019hkq}. 
Despite comprehensive searches, no new signals have been observed, 
leading to stringent upper limits on the product of  the two branching ratios, 
$\br(t\rightarrow b H^+) \br(H^+\rightarrow \tau^+\nu)$. 
For a given $\mch$, $\br(t\rightarrow b H^+)$ is determined solely by $\tan\beta$, whereas $\br(H^+\rightarrow \tau^+\nu)$ remains constant.
As a result, the observed upper bound on their product imposes a strict constraint on $\tan\beta$ for a given $\mch$, thereby limiting $\br(t\rightarrow b H^+)$.
For instance, at $M_{H^\pm} = 150\gev$ and $\tan\beta = 10$, we find that $\br(t\rightarrow b H^+)$ is only $7\times10^{-4}$.
Thus, top quark decay is not an effective production channel for $H^\pm$ at the LHC.

Given this constraint, alternative production channels for $H^\pm$ via the decay of heavier Higgs states, such as $H/A \rightarrow H^\pm W^{\mp(*)}$, have been extensively studied~\cite{Arhrib:2017wmo,Mondal:2021bxa,Arhrib:2021xmc,Biekotter:2023eil}.  The production of $H$ or $A$ occurs through gluon fusion ($gg\rightarrow H/A$) or associated production ($gg\rightarrow H \rightarrow A Z$, $q\bar{q}\rightarrow Z\rightarrow H Z$). However, these channels present two significant limitations. First, their signal rates strongly depend on model parameters $\mhha$ and $\tan\beta$. Second, as shown in \autoref{fig-mHc150-mHmA-tb}, the allowed parameter space forbids on-shell  decays for $H/A \rightarrow H^\pm W^{\mp}$, preventing us from using the $W$ boson mass constraint to effectively suppress backgrounds.

A more promising production mechanism is the pair production of charged Higgs bosons via the Drell-Yan process. This channel offers a straightforward and model-independent avenue for $H^\pm$ production, 
as the production cross section is exclusively determined by the mass of the charged Higgs boson. 
Therefore, we focus our analysis on the pair production channel 
for the $H^\pm \rightarrow t^* b$ decay mode.

\begin{figure}
%\vspace*{5cm}  
\centering
\includegraphics[width=0.48\textwidth]{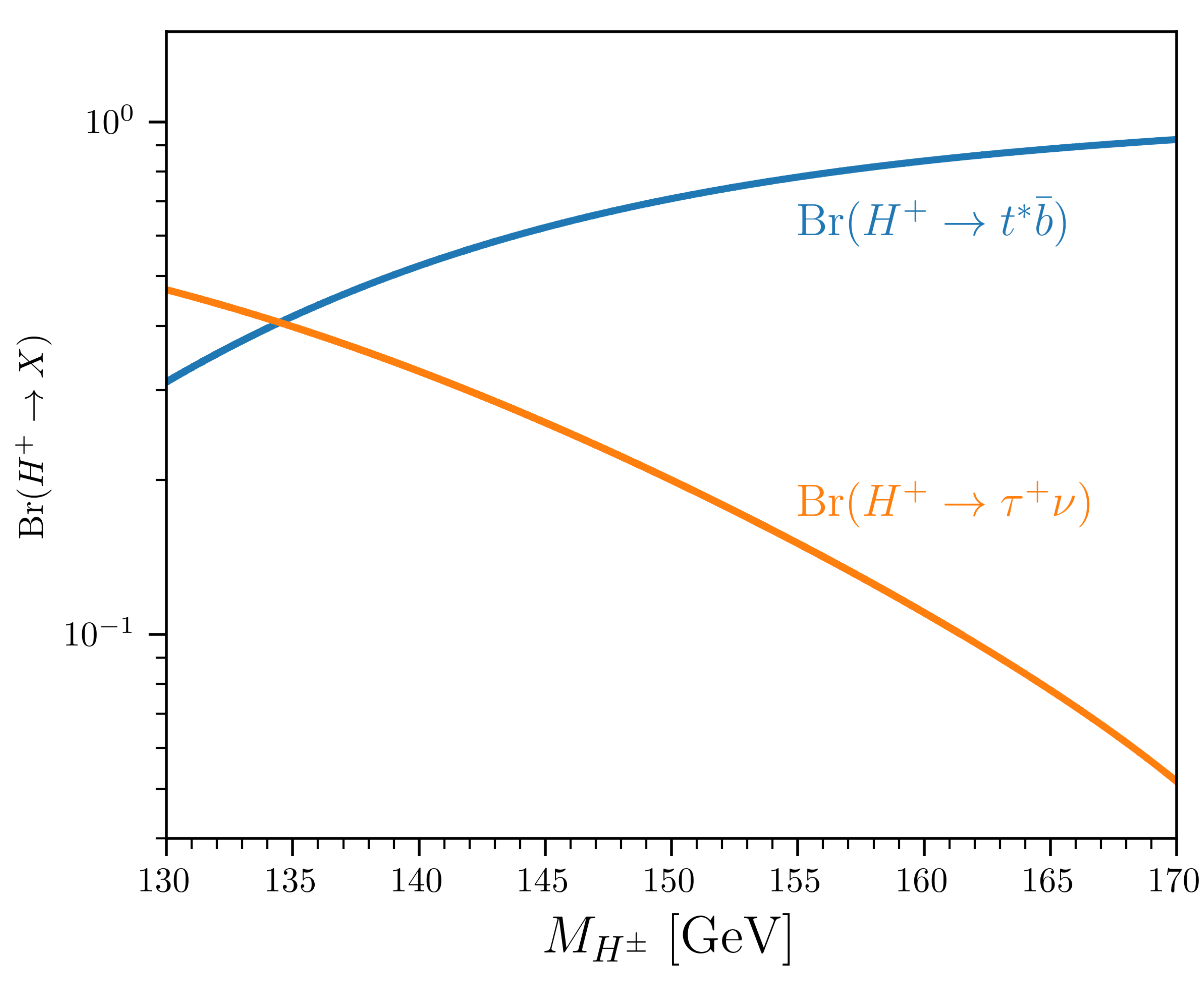}
\includegraphics[width=0.48\textwidth]{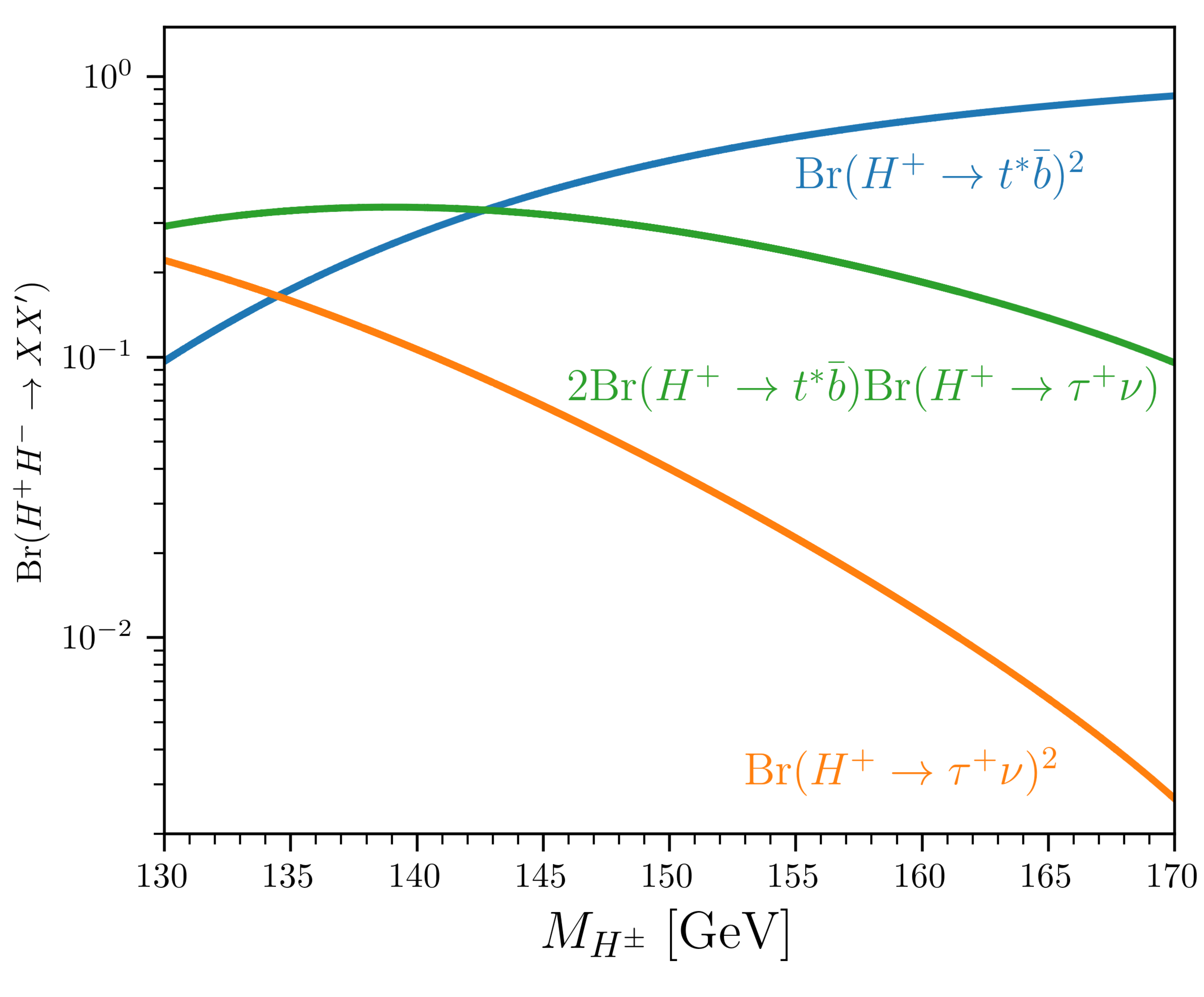}
\caption{Branching ratios of $\br(H^+ \rightarrow X)$ in the upper panel
and $\br(H^+ H^- \rightarrow XX')$ in the lower panel. We set $\mhh=\ma=200\gev$.
The branching ratios are independent of $\tan\beta$, as all Yukawa couplings of $H^\pm$ share a common factor of $1/\tan\beta$, making them dependent only on the fermion masses.}
\label{fig-BRs}
\end{figure}

Let us examine the decay patterns of the light charged Higgs boson in the mass range of 130--170 GeV. The dominant decay channels for $H^+$ are $t^* \bar{b}$ and $\tau^+ \nu$. \autoref{fig-BRs} shows their branching ratios as functions of $M_{H^\pm}$ for two scenarios: a single charged Higgs boson (upper panel) and a pair of charged Higgs bosons (lower panel). These branching ratios are independent of $\tb$, $m_{12}^2$, and $\mhha$. Our calculations, performed with {\small\sc 2HDMC}~\cite{Eriksson:2009ws}, include QCD radiative corrections up to order $\alpha_s^2$ in the $\overline{\text{MS}}$ scheme, incorporating running fermion masses in the Higgs couplings and leading logarithmic corrections to all orders at the renormalization scale $\mu_R = M_{H^\pm}$.

The decay mode $H^+ \rightarrow t^* \bar{b}$ maintains a significant branching ratio across the target $M_{H^\pm}$ range and becomes the dominant channel for $M_{H^\pm} \gtrsim 135$ GeV. For $M_{H^\pm} \lesssim 135$ GeV, however, $H^+ \rightarrow \tau^+ \nu$ takes precedence as the leading decay mode, with $H^+ \rightarrow t^* \bar{b}$ remaining subdominant.

In the pair production of charged Higgs bosons, the role of $H^+ \rightarrow t^* \bar{b}$ becomes even more pronounced. The process $H^+H^- \rightarrow t^* \bar{b} \, \bar{t}^* b$ dominates for $M_{H^\pm} \gtrsim 143$ GeV. 
Interestingly, the decay channel $H^+H^- \rightarrow t^* b \tau \nu$, which accounts for $H^+H^- \rightarrow t^* \bar{b} \tau^- \nu$ and its charge-conjugate process $H^+H^- \rightarrow  \bar{t}^* b \tau^+ \nu$, 
emerges as the most prominent mode for $M_{H^\pm} \lesssim 143$ GeV and remains the second most dominant for $M_{H^\pm} \gtrsim 143$ GeV, benefiting from a combinatorial factor of two.
In contrast, the $\tau\nu\tau\nu$ final state, which has been the focus of previous studies for the light $H^\pm$, exhibits significantly reduced branching ratios.

The results in \autoref{fig-BRs} strongly support investigating the off-shell $t^* b$ mode 
as a potential discovery channel for the light charged Higgs boson within the mass range 
in 130 to $170\gev$.
Since the $t^* b \, t^* b$ final state faces challenges from the combinatoric complications
and the larger QCD backgrounds at the LHC,
our investigation targets the following discovery channel
for the light charged Higgs boson in the mass range of 130 to 170 GeV:
\begin{align}
H^+ H^-  &\rightarrow  t^* b \tau\nu,
\end{align}
where $t^* b \tau\nu$ accounts for both charge-conjugate processes.

\section{$H^\pm \rightarrow t^* b$ at the HL-LHC and a 100 TeV $pp$ collider}
\label{sec-LHC}

In the preceding section, we identified the pair production of charged Higgs bosons, 
followed by $H^+H^- \rightarrow  t^* b \tau\nu$, 
as a key channel for probing the light charged Higgs boson in the mass range of $[130,170]\gev$. 
This section explores the discovery potential of the HL-LHC and a prospective 100 TeV $p p$ collider,
focusing on the case with $\mch=150\gev$. 
For the decay of the off-shell top quark, we consider the hadronic decay channel of the $W$ boson. Our signal process is summarized as:
\begin{equation}
\label{eq-LHC:final}
p p \rightarrow H^+ H^- \rightarrow [t^* (\rightarrow W b) b][ \tau \nu ] \rightarrow [j j b b][ \tau \nu],
\end{equation}
where $\tau=\tau^+,\tau^-$, $j$ denotes a light quark jet,
and the particles in a square bracket represent decay products originating from the same parent particle.

The resultant final state consists of two light jets, two $b$ jets, a tau lepton, and missing transverse energy. At the 14 TeV LHC, this final state can arise from several SM processes, including top quark pair production, $Wjjb\bar{b}$, $tbjj$ (single top production in the $t$-channel with an associated jet), $t\bar{t}W$, $t\bar{t}Z$, $t\bar{t}h_{\text{SM}}$, $tWZ$, and $t\bar{t}t\bar{t}$.

To evaluate their relative importance and identify the dominant background contributions, we compute the total cross sections for these processes at leading order using {\small\sc MadGraph5\_aMC@NLO}~\cite{Alwall:2011uj} version 3.5.0, with the {\small\sc NNPDF31\_nlo\_as\_0118} PDF set:
\begin{align}
\sigma(pp \to t\bar{t}) &= 530.2~\text{pb}, \\
\sigma(pp \to W^\pm jj b\bar{b}) &= 15.7~\text{pb}, \nonumber \\
\sigma(pp \to t\bar{t} b\bar{b} jj) &= 4.626~\text{pb}, \nonumber \\
\sigma(pp \to t\bar{t}W^\pm) &= 4.685 \times 10^{-1}~\text{pb}, \nonumber \\
\sigma(pp \to t\bar{t}Z) &= 6.103 \times 10^{-1}~\text{pb}, \nonumber \\
\sigma(pp \to t\bar{t}h_{\text{SM}}) &= 4.168 \times 10^{-1}~\text{pb}, \nonumber \\
\sigma(pp \to t\bar{t}W^\pm Z) &= 9.991 \times 10^{-2}~\text{pb}, \nonumber \\
\sigma(pp \to t\bar{t}t\bar{t}) &= 5.135 \times 10^{-3}~\text{pb}, \nonumber
\end{align}
where $\ttt = t,\bar{t}$ and $\bbb = b,\bar{b}$. For processes involving jets and/or $b$ jets, we imposed the kinematic cuts $p_T^{j,b} > 25~\text{GeV}$, $|\eta_{j,b}| < 2.5$, and $\Delta R_{j_i, j_j} > 0.4$ for $j_{i,j} = j, b$.
For cross-validation, 
we additionally used \textsc{CalcHEP}~\cite{Belyaev:2012qa} to calculate the cross sections
for  several  channels such as \( pp\to t\bar{t}W^\pm \) and \(  pp\to Wjjb\bar{b} \). The resulting parton-level cross sections agree within expected statistical and systematic uncertainties, confirming the internal consistency of our simulation setup.

As evident from the computed cross sections, top quark pair production is by far the most significant background, with a rate exceeding those of all other processes by more than an order of magnitude. Accordingly, our analysis primarily focuses on this dominant background channel:
\begin{equation}
\label{eq-final}
pp \rightarrow t\bar{t} \rightarrow [bW^+][\bar{b}W^-] \rightarrow [bjj][b\tau\nu],
\end{equation}
where $[bjj][b\tau\nu]$ implicitly includes all possible charge-conjugate combinations.

To quantitatively assess both the signal and the dominant $t\bar{t}$ background, we carried out a detailed parton-level simulation at the 14 TeV LHC. Cross sections were computed using {\small\sc MadGraph5\_aMC@NLO} version 3.5.0 with the {\small\sc NNPDF31\_nlo\_as\_0118} PDF set. The renormalization and factorization scales were dynamically set to $\mu_{R} = \mu_{F} = \sum_i \sqrt{p_{T,i}^2 + m_i^2}$, where the sum runs over all final-state particles. In total, $3.2 \times 10^6$ signal events and $1.2 \times 10^7$ background events were generated for further analysis.

To ensure accurate event modeling, we went beyond the leading-order approximation: the signal was simulated at next-to-leading order (NLO), while the $t\bar{t}$ background was corrected to approximate next-to-next-to-next-to-leading order (aN$^3$LO) using $K$-factors based on state-of-the-art theoretical predictions~\cite{Nason:1987xz,Beenakker:1990maa,Czakon:2011xx,Kidonakis:2019yji,Kidonakis:2022hfa,Kim:2024ppt}. The resulting parton-level cross sections for the final state in \autoref{eq-final} are $1.593 \times 10~\fb$ for the signal and $1.028 \times 10^2~\pb$ for the background.

Parton showering and hadronization were performed using {\small\sc Pythia} version 8.309~\cite{Bierlich:2022pfr}. For the detector-level analysis, a fast detector response simulation was employed with {\small\sc Delphes}~\cite{deFavereau:2013fsa} using the high-luminosity LHC card
\texttt{delphes\_card\_HLLHC.tcl}. Jet clustering was conducted with {\small\sc FastJet} version 3.3.4~\cite{Cacciari:2011ma} using the anti-$k_T$ algorithm with a jet radius of $R=0.4$.

Accurate identification of the final state in \autoref{eq-LHC:final} critically depends on $b$-tagging 
and $\tau$-tagging procedures. A jet is designated as a $b$ jet if a $B$ hadron with $p_T > 5\gev$ is detected within a $\Delta R = 0.3$ radius of the jet. Candidate $b$ jets are required to meet the threshold of $p_T > 25\gev$ and $|\eta| < 2.5$, after which the $b$-tagging efficiency is applied. Charm and other light quark jets can be mistagged as $b$ jets. 
The efficiencies for $b$ tagging and mistagging depend on the jet's kinematics and are approximately~\cite{ATL-PHYS-PUB-2016-026,ATL-PHYS-PUB-2017-001}:
\bea
\label{eq-b-tagging}
P_{b\rightarrow b} \simeq 75\%,\quad P_{c\rightarrow b} \simeq 10\%,\quad P_{j \rightarrow b} \simeq 1\%.
\eea

Identification of the tau lepton is feasible when it decays hadronically, denoted as $\tauh$, marked by a collimated jet with a sparse number of hadrons~\cite{CMS:2007sch,Bagliesi:2007qx,CMS:2018jrd}. 
The {\small\sc Delphes} default settings for $\tau$ tagging and mistagging efficiencies were applied,
which are approximately
\bea
\label{eq-tau-tagging}
P_{\tau \rightarrow \tauh} \simeq 60\%,
\quad P_{e\rightarrow \tauh} \simeq 0.5\%, \quad
P_{j \rightarrow \tauh} \simeq 1\%.
\eea

After completing the detector simulation, the following basic selection criteria are imposed:
\begin{itemize}
\item $N_j \ge 2$, $N_b \ge 2$, and $N_\tauh \ge 1$, where $j$, $b$, and $\tauh$ satisfy $p_T > 25$ GeV and $|\eta| < 2.5$;
\item $\met > 25$ GeV.
\end{itemize}
The requirement of a minimum missing transverse energy ($\met$) is essential since our signal includes a neutrino from the decay $H^\pm \rightarrow \tau\nu$.

Despite the relatively loose selection criteria, the basic selection results in
an exceedingly low acceptance rate for the signal, approximately 1\%. 
In contrast, the background acceptance rate is substantially higher,
around 2.7\%. With the suppressed signal cross section after the basic selection being only $1.852 \times 10^{-1} \fb$, even the total luminosity of $3 \iab$ at the HL-LHC would result in merely a few hundred signal events. This scarcity of signal events severely limits the ability to devise an effective strategy using kinematic cuts to disentangle the signal from the backgrounds, irrespective of their efficiency.

\begin{figure*}[t]
    \centering
    \includegraphics[width=\textwidth]{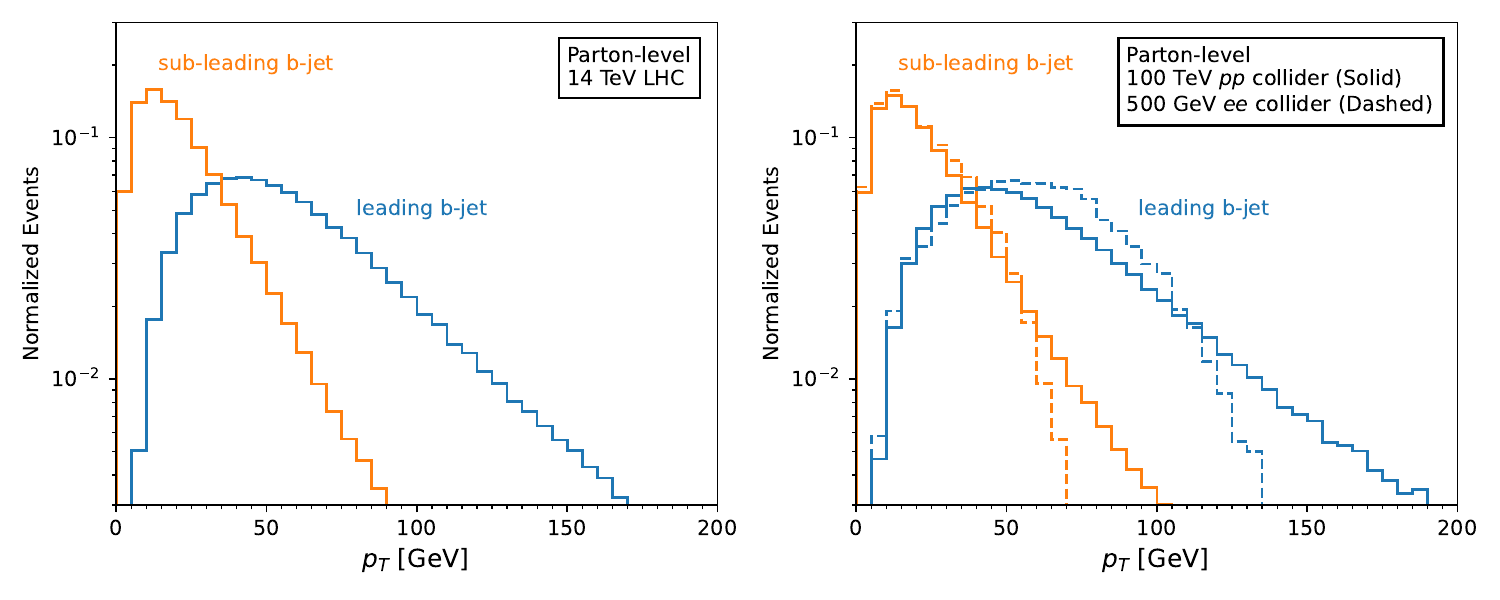}
  \caption{Normalized distributions of transverse momenta for the leading and subleading $b$-jets at the parton level in the signal process $pp \rightarrow H^+ H^- \rightarrow t^* b\tau\nu \rightarrow jjbb\tau\nu$, with $\mch=150\gev$. The left panel shows the distributions at the 14 TeV LHC, while the right panel presents the results for a 100 TeV $pp$ collider (solid lines) and a 500 GeV $e^+ e^-$ collider (dashed lines).}
  \label{fig:parton:bjet:pt:LHC:100TeV}
\end{figure*}

A primary factor contributing to the low signal acceptance after basic selection 
is the low transverse momentum (softness) of the $b$ jets originating from the decay $\ch\rightarrow t^* b$.
This softness is due to the small mass difference between $\mch$ and $m_t$, 
which is crucial for ensuring a substantial branching ratio for the $\ch\rightarrow t^* b$ decay mode, 
as depicted in \autoref{fig-BRs}. 
Additionally, the $b$ quark from the decay of the off-shell top quark, $t^* \rightarrow W b$, 
also exhibits lower transverse momentum compared to its on-shell counterpart. 
The challenge with these soft $b$ jets is that 
the majority of the subleading $b$ jets fail to meet the minimum transverse momentum 
threshold ($p_T > 25\gev$) required for $b$ jet clustering.\footnote{Current $b$-tagging methods at ATLAS and CMS typically require a minimum jet $p_T$ threshold of about 20 GeV~\cite{ATLAS:2019bwq,CMS:2017wtu}, while the FCC-hh studies assume thresholds above 30 GeV~\cite{Mangano:2022ukr}. Although $b$-tagging algorithms are continually improving, lowering this threshold remains particularly challenging. Soft $b$-jets have shorter flight distances, making them difficult to distinguish from light jets. Furthermore, they suffer from lower track reconstruction efficiency, complicating their identification. As demonstrated in Ref.~\cite{CMS:2017wtu}, decreasing the $p_T$ threshold further leads to increased background contamination from misidentified light jets, highlighting an intrinsic limitation of $b$-tagging performance at low $p_T$.}

\autoref{fig:parton:bjet:pt:LHC:100TeV} presents the parton-level $p_T$ distributions of the leading and subleading $b$ jets, ordered by descending $p_T$, for the signal process. The left panel shows results at the 14 TeV LHC, while the right panel compares results from a 100 TeV $pp$ collider (solid lines) and a 500 GeV $e^+ e^-$ collider (dashed lines). At all three collider configurations, approximately 70\% of the subleading $b$ jets fail to exceed the $p_T > 25\gev$ threshold. This threshold is essential for both $b$-jet tagging and jet clustering algorithms, serving to reduce low-energy noise, suppress backgrounds, and improve computational efficiency. Furthermore, lowering this threshold would be counterproductive, as it would increase background acceptance more than signal acceptance.

Even at the higher collision energy of 100 TeV (solid lines, right panel), both the leading and subleading $b$ jets remain similarly soft as those at the HL-LHC. The $b$ jets do not gain substantial transverse boosts, despite the significant increase in c.m.~energy. This occurs because the steep decline of the parton distribution functions at large momentum fractions  prevents parton-level collisions at hadron colliders from fully utilizing the available beam energy.

One might wonder whether future $e^+e^-$ colliders could provide sufficient boost,
which is crucially determined by their c.m.~energies. Circular colliders such as FCC-ee~\cite{TLEPDesignStudyWorkingGroup:2013myl,FCC:2018evy} and CEPC~\cite{An:2018dwb} are constrained by synchrotron radiation, limiting them to $\sqrt{s} \leq 365$ GeV~\cite{Naseem:2019iua}. The International Linear Collider (ILC)~\cite{Moortgat-Pick:2015lbx,LCCPhysicsWorkingGroup:2019fvj,Bhattacharya:2023mjr} is 
planned for maximum c.m.~energies of 500 GeV, while the Compact Linear Collider (CLIC)~\cite{Aicheler:2012bya} has initial\footnote{Although the CLIC aims for a final goal of $\sqrt{s}=3$ TeV, achieving this would require a 50 km tunnel, making it a distant prospect.} c.m.~energies of 380 GeV.

Since 500 GeV represents the highest achievable c.m. energy for near-future $e^+e^-$ colliders, we examine the $p_T$ distributions of $b$ jets in our signal process $e^+ e^- \to H^+ H^- \\ \to bbjj\tau\nu$ at $\sqrt{s}=500\gev$ (dashed lines in the right panel of \autoref{fig:parton:bjet:pt:LHC:100TeV}). These distributions closely resemble those at a 100 TeV $pp$ collider, demonstrating that even at $e^+e^-$ colliders, the $b$ jets—especially the subleading one—remain insufficiently boosted. While $e^+e^-$ colliders offer excellent physics opportunities and discovery potential for various new particles, their currently proposed energy range falls short of what is needed to effectively probe our signal process.

Given the small signal yield after basic selection, we focus on developing kinematic cuts that maximize signal retention while effectively reducing backgrounds. With this goal in mind, we examined various kinematic distributions and identified key variables that could discriminate the signal from the background.

One set of crucial discriminating variables are the angular separations,
defined  by $\Delta R \equiv \sqrt{(\Delta \eta)^2 + (\Delta \phi)^2}$, among $b$ jets and light jets. These variables are efficient because the signal channel $pp \rightarrow H^+ H^- \rightarrow [jjbb][\tau\nu]$ 
results in a smaller angular separation within the $[jjbb]$ grouping.  In contrast, the background channel $pp \rightarrow t\bar{t} \rightarrow [bjj][b\tau\nu]$ forms a back-to-back topology between $[b jj ]$ and $[b\tau\nu]$, leading to a larger $\Delta R$ between two $b$ jets
as well as  a larger $\Delta R$ between $j$ and the $b$ in the $[b\tau\nu]$ system.

\begin{figure*}[!t]
  \centering
  \includegraphics[width=\textwidth]{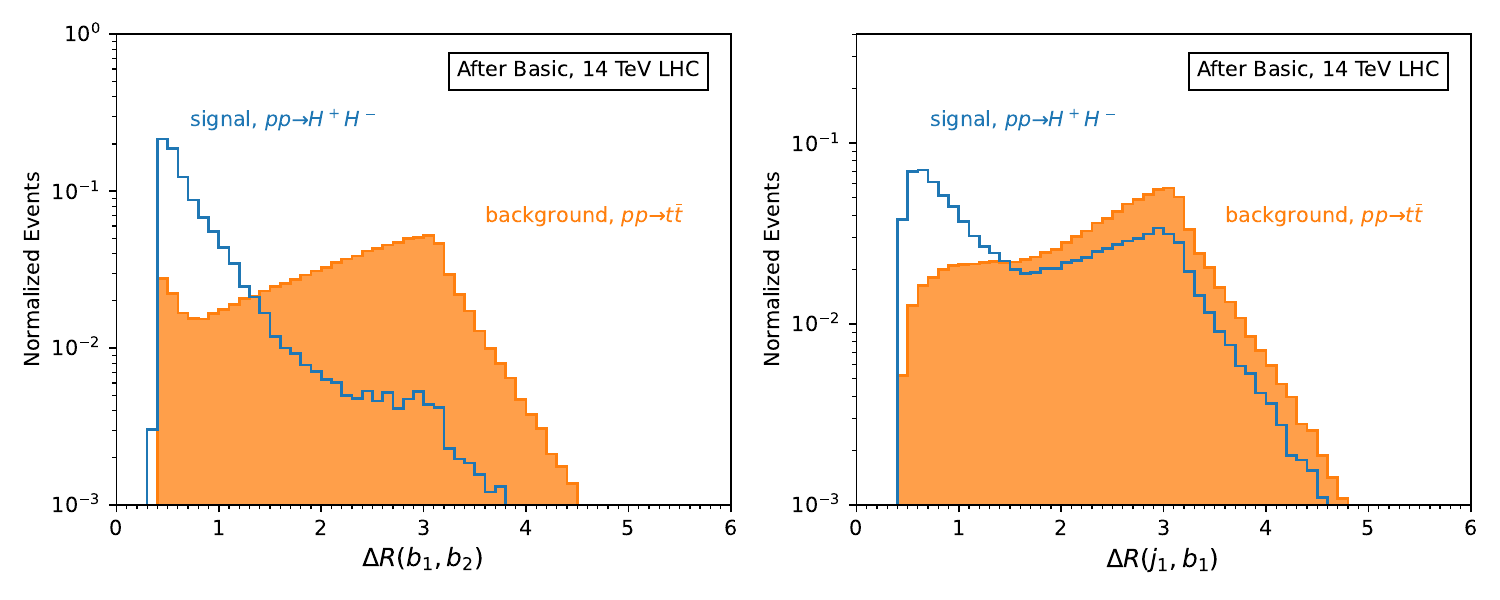}
  \vspace{-0.7cm}
  \caption{Normalized distributions of $\Delta R(b_1, b_2)$ (left) and $\Delta R(j_1, b_1)$ (right) 
  after applying the basic selection criteria outlined in the main text. 
  The signal results with $\mch=150\gev$ are presented by blue solid lines
  while the $t\bar{t}$ background results are by orange histograms, respectively. 
}
  \label{fig-tbtaunu-LHC-DR1}
\end{figure*}

In \autoref{fig-tbtaunu-LHC-DR1}, we present normalized distributions of two representative angular separations: $\Delta R(b_1, b_2)$ in the left panel and $\Delta R(j_1, b_1)$ in the right panel. For the signal $pp \rightarrow H^+ H^- \rightarrow [jjbb][\tau\nu]$, the $\Delta R(b_1, b_2)$ distribution peaks near 0.4, indicating the close proximity of the two $b$ jets originating from the same parent particle $H^\pm$.
In contrast, the background $pp \rightarrow t\bar{t} \rightarrow [bjj][b\tau\nu]$ exhibits a broader $\Delta R(b_1, b_2)$ distribution with a dominant peak near 3. This reflects the back-to-back motion of two $b$ jets, each originating from a different parent top quark.
Moreover, the $\Delta R(j_1, b_1)$ distributions further highlight differences between the signal and background. The signal's $\Delta R(j_1, b_1)$ distribution peaks at a lower value of approximately 0.8, whereas the background peaks at higher values near 3.

Based on these distinct features in the angular separation distributions, we impose the following $\Delta R$ cuts to suppress the background while retaining a significant fraction of the signal events:
\bea 
\label{eq-DR:cut} 
\Delta R (b_1, b_2) < 0.8, \, \Delta R (j_i, b_j) < 1.5  \, \text{ for } i, j = 1, 2. 
\eea

Other crucial discriminating variables pertain to the reconstruction of the charged Higgs boson mass $\mch$. In the signal process $H^+ H^- \rightarrow [bbjj][\tau\nu]$, $\mch$ can be measured in two complementary ways: through the invariant mass of the $[bbjj]$ system and the transverse mass derived from the $[\tau\nu]$ system.
The transverse mass $M_T(X)$ is a useful variable defined for a visible particle $X$ and the missing transverse energy $\vec{E}_T^{\rm miss} = -\sum_i \vec{p}_T^{, i}$, where $i$ covers all observed particles~\cite{Smith:1983aa,Han:2005mu}:
\bea
\label{eq-mtch}
M_T(X)= 
\sqrt{m_{X}^2
+ 2\left[
E_T^{X} E_T^{\rm miss} - \vec{p}_T^{\,X} \cdot\vec{E}_T^{\rm \, miss}
\right]}
,
\eea
where $E_T^X = \sqrt{m_X^2 + (p_T^X)^2}$. 

For the signal process $H^\pm \rightarrow\tau\nu$,
with the visible particle being the tau lepton, 
$M_T(\tau)$ is expected to peak at the charged Higgs boson mass $\mch$.
Since both the invariant mass $M_{bbjj}$ and the transverse mass $M_T(\tau)$ should reconstruct 
the same $\mch$ for the signal events, 
we define an asymmetry variable $\mathcal{A}_M$ to quantify the difference between these two mass observables:
\begin{eqnarray}
\label{eq-asymmetry}
\mathcal{A}_M= \left| \frac{M_{bbjj} - M_T(\tau)}{M_{bbjj} +  M_T(\tau)} \right| .
\end{eqnarray}
By imposing an appropriate upper bound on $\mathcal{A}_M$, we can efficiently separate the signal  from the background.

\begin{figure*}[!t]
  \centering
  \includegraphics[width=\textwidth]{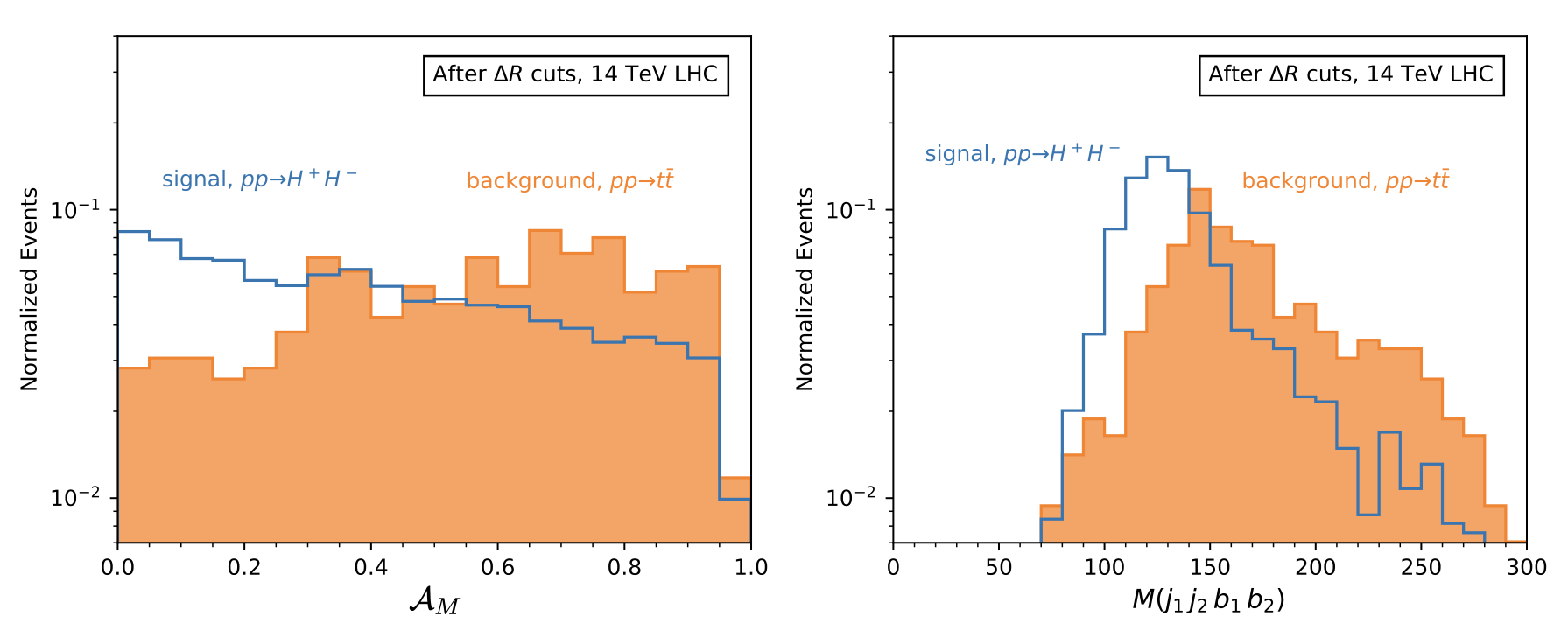}
  \vspace{-0.7cm}
  \caption{Normalized distributions of the mass asymmetry $\mathcal{A}_M$ (left) and the invariant mass $M(j_1j_2b_1b_2)$ (right), plotted after implementing $\Delta R (b_1,b_2)< 0.8$ and $\Delta R (j_i,b_j)< 1.5$ for $i,j=1,2$. The definition of $\mathcal{A}_M$ is provided in the main text. The blue curves represent the signal with $\mch=150\gev$, while the orange histograms depict the $t\bar{t}$ background.
}
  \label{fig-tbtaunu-LHC-AM}
\end{figure*}

In \autoref{fig-tbtaunu-LHC-AM}, we present the normalized distributions of the asymmetry variable $\mathcal{A}_M$ (left panel) and the invariant mass $M(j_1 j_2 b_1 b_2)$ (right panel) for the signal with $\mch=150\gev$ (blue lines) and the $t\bar{t}$ background (orange histograms), after imposing the $\Delta R$ cuts defined in \autoref{eq-DR:cut}. Two distinct features are immediately evident. First, as expected, the $\mathcal{A}_M$ distribution for the signal peaks sharply near $\mathcal{A}_M \simeq 0$, whereas the background distribution is concentrated at higher values, typically above 0.2. Thus, applying an upper bound on $\mathcal{A}_M$ will effectively enhance signal-background separation.

The second notable feature is observed in the invariant mass distribution $M(j_1j_2b_1b_2)$. 
For the signal, it exhibits a resonance peak around 130 GeV, 
which is lower than the true charged Higgs boson mass of 150 GeV. 
This discrepancy is attributed to several factors, 
including neutrinos in $B$ meson decays, imperfect jet energy resolution, 
and gluon radiation not fully captured within the jet clustering cone. 
Despite this shift, the resonance peak in the $M(j_1j_2b_1b_2)$ distribution provides a distinct signature for the signal process.

\begin{table*}[h]
  \centering
  \setlength{\tabcolsep}{10pt}
  \renewcommand{\arraystretch}{1.3}
  \begin{tabular}{|l||c|c||c|}
  \hline
\multicolumn{4}{|c|}{Cut-flow for $pp\rightarrow H^+ H^- \rightarrow t^* b \tau\nu\rightarrow jjbb\tau\nu$ at the 14 TeV LHC}
\\ \hline
   & $\sigma_{\rm sg}$ [fb] & $\sigma_{\rm bg}$ [fb] & $\mathcal{S}^{10\%}_{3 {\rm ab}^{-1}}$ \\ \hline
  Basic Selection& $1.852 \times 10^{-1}$ & $2.814 \times 10^{3}$ & $6.578 \times 10^{-4}$ \\ \hline
  $\Delta R (b_1,b_2)<0.8$  & $1.139 \times 10^{-1}$ & $2.326 \times 10^{2}$ & $4.897 \times 10^{-3}$ \\ \hline
  $\Delta R (j_1,b_1)<1.5 $  & $5.319 \times 10^{-2}$ & $3.198 \times 10^{1}$ & $1.661 \times 10^{-2}$ \\ \hline
  $\Delta R (j_1,b_2)<1.5 $  & $4.860 \times 10^{-2}$ & $2.460 \times 10^{1}$ & $1.973 \times 10^{-2}$ \\ \hline
  $\Delta R (j_2,b_1)<1.5 $  & $1.989 \times 10^{-2}$ & $4.648$ & $4.257 \times 10^{-2}$ \\ \hline
  $\Delta R (j_2,b_2)<1.5 $  & $1.707 \times 10^{-2}$ & $3.625$ & $4.680 \times 10^{-2}$ \\ \hline
  $\mathcal{A}_M < 0.2$  & $5.073 \times 10^{-3}$ & $4.179 \times 10^{-1}$ & $1.164 \times 10^{-1}$ \\ \hline
  $M(j_1j_2b_1b_2)<m_t$  & $4.779 \times 10^{-3}$ & $2.644 \times 10^{-1}$ & $1.694\times 10^{-1}$ \\ \hline
  $M_T(\tauh)<m_t $  & $4.650 \times 10^{-3}$ & $2.303 \times 10^{-1}$ & $1.875 \times 10^{-1}$ \\ \hline
  \end{tabular}
  \caption{Cut-flow for the signal process $p p \rightarrow H^+ H^- \rightarrow t^* b \tau \nu \rightarrow j j b b \tau \nu$ with $\mch=150\gev$
  and the dominant background process $p p \rightarrow t \bar{t}  \rightarrow j j b b \tau \nu$. 
  The significance is calculated considering a 10\% background uncertainty 
  and an integrated luminosity of $3 \iab$.}
  \label{tab:cutflow_HLLHC}
\end{table*}

To evaluate the discovery potential for this channel, 
we present in Table \ref{tab:cutflow_HLLHC} 
the cut-flow for the signal process 
$pp\rightarrow H^+ H^- \rightarrow t^* b \tau\nu\rightarrow jjbb\tau\nu$ at the 14 TeV LHC. 
The table shows the cross sections of the signal and background after each cut, 
as well as the significance considering a 10\% background uncertainty for an integrated luminosity of $3 \iab$. 
The significance is defined as~\cite{Cowan:2010js}:
\bea
\label{eq-significance:nbg}
\mathcal{S} &=& 
\left[2(\nsg + \nbg) \log\left(\frac{(\nsg + \nbg)(\nbg + \dbg^2)}{\nbg^2 + (\nsg + \nbg)\dbg^2} \right)  \right.
\\ \nn && \left.
- 
\frac{2 \nbg^2}{\delta_b^2} \log\left(1 + \frac{\dbg^2 \nsg}{\nbg (\nbg + \dbg^2)}\right)\right]^{1/2},
\eea
where $\nsg$ denotes the number of signal events, 
$\nbg$ the number of background events, and $\dbg = \Dbg  \nbg$ the background uncertainty yield.

Despite applying the key kinematic cuts, the final significance reaches only 0.19. 
While our final selection cut boosts the signal significance by approximately 300-fold 
relative to the basic selection step, 
the significance remains substantially below a detectable level. 
Moreover, the signal cross section after the final selection, approximately $4.65 \ab$, precludes any further refinements through cuts.
This cut-based analysis highlights the challenges in probing the charged Higgs boson 
through the $H^+ H^- \rightarrow t^* b \tau\nu$ channel at the HL-LHC using kinematic cuts alone, 
due to the extremely small signal cross section and overwhelming backgrounds.

To assess if the BDT method can enhance sensitivity 
to the $t^* b$ decay mode of the light charged Higgs boson, 
we employed the Extreme Gradient Boosting (XGBoost) package~\cite{Chen:2016btl}. 
XGBoost has seen increasing use in the particle physics community for a variety of analyses, 
including studies on the SM Higgs boson~\cite{ATLAS:2017ztq,CMS:2020tkr,ATLAS:2021ifb,CMS:2020cga,CMS:2021sdq}, 
dark matter~\cite{ATLAS:2021jbf}, vectorlike quarks~\cite{Dasgupta:2021fzw}, 
a composite pseudoscalar~\cite{Cornell:2020usb}, 
and innovative strategies for faster event generation~\cite{Bishara:2019iwh}.
In our study, XGBoost is used as a binary classifier, 
aimed at more effectively distinguishing between signal and background events.

We initialized the XGBoost classifier with the objective set to \texttt{binary} and the evaluation metric set to \texttt{logloss}.  
The learning rate was configured at 0.1. 
For training the model, we generated $3.7\times 10^4$ signal events 
and $3.3\times 10^5$ background events, all of which met the basic selection criteria. 
We divided the dataset into three parts: 
50\% for training, 20\% for validating the algorithm, and 30\% for testing. 
As inputs for the XGBoost model, we used the following 51 variables:\footnote{A large number of input parameters does not pose a concern in XGBoost, as its tree-based structure inherently selects only the most relevant features. This prevents overfitting and ensures that unnecessary or correlated inputs do not impact model performance.}
\begin{enumerate}
  \item Angular distance: 
  \begin{align}\nn 
&\Delta R (b_1, b_2) ,\quad
  \Delta R (b_1, j_1j_2)  ,\quad
  \Delta R (b_1, j_1j_2b_2),
  \\ \nn &
  \Delta R (b_1, \tauh),\quad
  \Delta R (b_1\tauh, j_1j_2), \quad \Delta R (b_2, j_1j_2),
 \end{align}
   \begin{align}\nn  &
\Delta R (b_2, j_1j_2b_1), \quad \Delta R (b_2, \tauh), \quad \Delta R (b_2\tauh, j_1j_2),
\\ \nn & \Delta R (j_1, b_1),  \quad \Delta R (j_1, b_2), \quad  \Delta R (j_1, j_2),
\\ \nn &
\Delta R (j_1, \tauh), \quad \Delta R (j_2, b_1), \quad \Delta R (j_2, b_2),
  \\ \nn &
 \Delta R (j_2, \tauh), \quad  \Delta R (\tauh, j_1j_2), \quad \Delta R (\tauh, j_1j_2b_1),
   \\ \nn &
\Delta R (\tauh, j_1j_2), \quad    \Delta R (\tauh, j_1j_2b_1b_2), \quad \Delta R (\tauh, j_1j_2b_2).
  \end{align}
  \item Invariant mass: \\      $M(j_1j_2)$, $M(j_1j_2b_1)$, $M(j_1j_2b_1b_2)$, $M(j_1j_2b_2)$.
  \item Transverse mass:\\       $M_T(b_1\tauh)$, $M_T(b_2\tauh)$, $M_T(\tauh)$.
  \item Four momentum of $\left[ p_T, \eta, \phi, m \right]$ for each of $j_1$, $j_2$, $b_1$, $b_2$, $\tauh$.
  \item Missing transverse energy and Missing energy azimuthal angle: \\
      $E_T^{\rm miss}$, $\phi^{\rm miss}$.
  \item Mass asymmetry:       $\mathcal{A}_M$.
\end{enumerate}
Here, a symbol representing multiple particles denotes a composite system formed by its constituent particles. For example, in $\Delta R (\tauh, j_1j_2)$, the term $j_1j_2$ represents a system composed of $j_1$ and $j_2$, where the four-momentum of this composite system is defined as the vector sum of the individual momenta of $j_1$ and $j_2$. Thus, $\Delta R (\tauh, j_1j_2)$ corresponds to the angular distance between $\tauh$ and the combined system of $j_1$ and $j_2$. This convention is consistently applied throughout our feature set.

\begin{figure}[!t]
  \centering
  \includegraphics[width=0.49\textwidth]{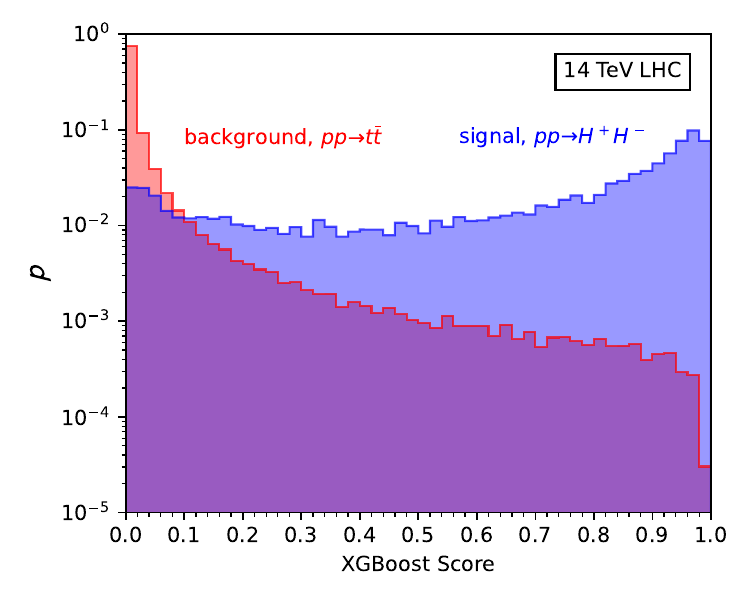}
  \caption{Normalized distributions of the signal (blue)  and backgrounds (red)  against the BDT score,
  based on the testing dataset.
}
  \label{fig-bdt}
\end{figure}

In \autoref{fig-bdt}, we illustrate the distributions of BDT scores 
for both the signal (blue) and the background (red). 
These results are derived exclusively from the testing dataset, 
which the model did not encounter during its training and validation phases. 
The BDT score distributions reveal a discernible separation between the signal and background. 
Applying a threshold of 0.98 for the XGBoost score 
results in cross sections of $1.42 \times 10^{-2} \fb$ for the signal 
and $8.53 \times 10^{-2} \fb$ for the background. 
Considering a 10\% background uncertainty and an integrated luminosity of $3 \iab$, 
the signal significance reaches 1.35.

Although the signal significance in the BDT analysis shows a roughly sevenfold increase 
compared to the significance in the cut-based analysis, it remains below the threshold for a confident detection. 
This challenge is primarily due to the soft $b$ jets in the signal events,
which fail to satisfy the basic selection. 

To explore whether a 100 TeV $pp$ collider could offer higher discovery potential 
for the $H^\pm \to t^* b$ signal, 
we conducted signal-to-background analyses for the same process, 
$pp \to H^+ H^- \to t^* b \tau\nu \to bbjj \tau\nu$, using the \textsc{Delphes} card \texttt{FCChh.tcl}. 
Unfortunately, the issue of excessively soft $b$-jets persists even at a 100 TeV $pp$ collider, 
failing to yield a significance above the detection threshold.

Our cut-based analysis, employing sequential kinematic cuts of $N_\tauh \geq 1$ (with $p_T^\tauh >60\gev$), 
$N_b \geq 2$, $N_j \geq 2$, $\Delta R (b_1,b_2)<1.5$, $\met\geq 100\gev$, 
$\mathcal{A}_M<0.7$, and $M(j_1j_2b_1b_2)<200\gev$, 
results in a significance of only 0.38 at a 100 TeV $pp$ collider. 
Moreover, the BDT analysis underperforms compared to the HL-LHC, 
with the signal significance reaching merely about 0.75.

In conclusion, high-energy hadron colliders, such as the HL-LHC and a prospective 100 TeV $pp$ collider, cannot effectively probe the $t^* b$ decay mode of the light charged Higgs boson due to the inherent softness of the $b$-jets in this channel. This limitation highlights the need for an alternative collider to investigate this particular decay mode of the light charged Higgs boson.

\section{$H^\pm \rightarrow t^* b$ at a Multi-TeV Muon Collider}
\label{sec-MuC}

Our analysis in the previous section showed that the signal process $pp \rightarrow H^+H^- \rightarrow t^* b \tau\nu$ yields insufficient significance at hadron colliders, primarily because the $b$ jets from $H^\pm \rightarrow t^* b$ decay are too soft for effective reconstruction. Even at a 100 TeV $pp$ collider, where only a fraction of the beam energy contributes to the parton-parton collision, the $b$ jets remain insufficiently boosted. This limitation motivates us to consider a multi-TeV MuC, where the full beam energy is available in collisions between fundamental particles. We examine two configurations~\cite{Han:2021udl,AlAli:2021let}: $\sqrt{s}=3\tev$ with an integrated luminosity of $1\iab$ and $\sqrt{s}=10\tev$ with $10\iab$.

For the production of charged Higgs bosons,
we consider their pair production with the following final state:
\bea
\label{eq-final:state}
H^+H^-\rightarrow [t^* b] [\tau\nu] \rightarrow [b b jj][ \tau\nu].
\eea 
At the MuC, there are three different production channels relevant to this final state:
\begin{align}
\label{eq-signal-DY-muc}
\mu^- \mu^+ &\rightarrow H^+ H^-,
\\ \label{eq-signal-CC-muc}
\mu^+\mu^- &\rightarrow H^+ H^-\nu \bar{\nu},
\\ \label{eq-signal-NC-muc}
\mu^+\mu^- &\rightarrow H^+ H^- \muf^+\muf^-,
\end{align}
where $\nu$ and $\bar{\nu}$ include all three neutrino flavors,
and $\muf$ denotes a forward muon with $|\eta|>2.5$.

The first process in \autoref{eq-signal-DY-muc} represents the Drell-Yan process mediated by the $Z$ boson and photon. The second process in \autoref{eq-signal-CC-muc} involves $H^\pm$ pair production with two additional neutrinos. Since these neutrinos appear as missing transverse energy in the detector, this process creates a signature identical to our target process with a single neutrino in \autoref{eq-final:state}. For this process, there are multiple contributing Feynman diagrams, with vector boson scattering (VBS) processes through $t$-channel $H$ and $A$ mediators being particularly important. We provide a detailed discussion of this process in \autoref{appendix}.

The third process in \autoref{eq-signal-NC-muc} describes $H^\pm$ pair production with two forward muons. In conventional multi-TeV MuC detector designs, these forward muons cannot be detected due to tungsten nozzles\footnote{Recently, the MuC community has begun integrating forward muon detectors into their designs, as energetic forward muons can penetrate the tungsten nozzles. These detectors are expected to play a crucial role in probing various BSM models~\cite{Accettura:2023ked,Ruhdorfer:2023uea,Forslund:2023reu,Bandyopadhyay:2024plc}.} that shield the detector from BIB particles~\cite{Collamati:2021sbv,Ally:2022rgk}, limiting pseudorapidity coverage to $|\eta|<2.5$. This neutral-current VBS process proceeds through several channels: $ZZ\to h^* \to H^+ H^-$, the quartic vertex $Z/\gamma$-$Z/\gamma$-$H^+$-$H^-$, and $t$-channel exchanges of $H$ and $A$.

\begin{figure}[!t]
  \centering
   \includegraphics[width=0.47\textwidth]{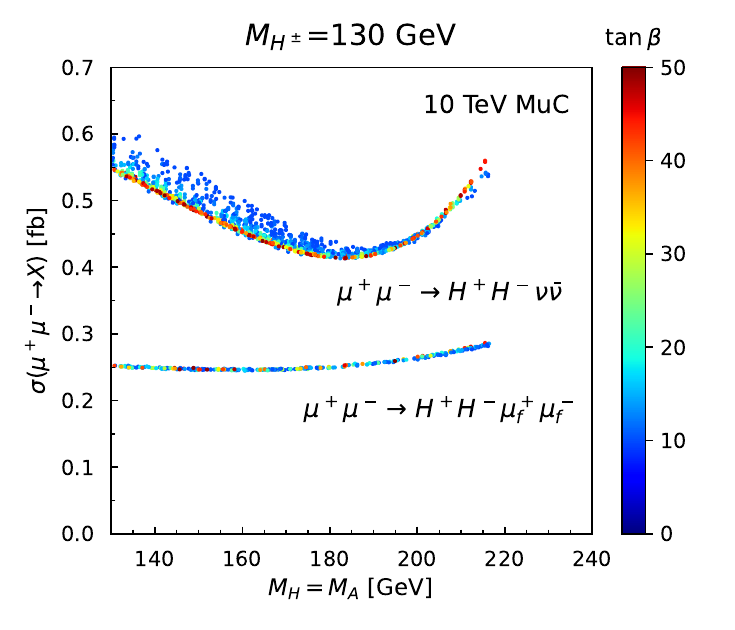}
  \includegraphics[width=0.47\textwidth]{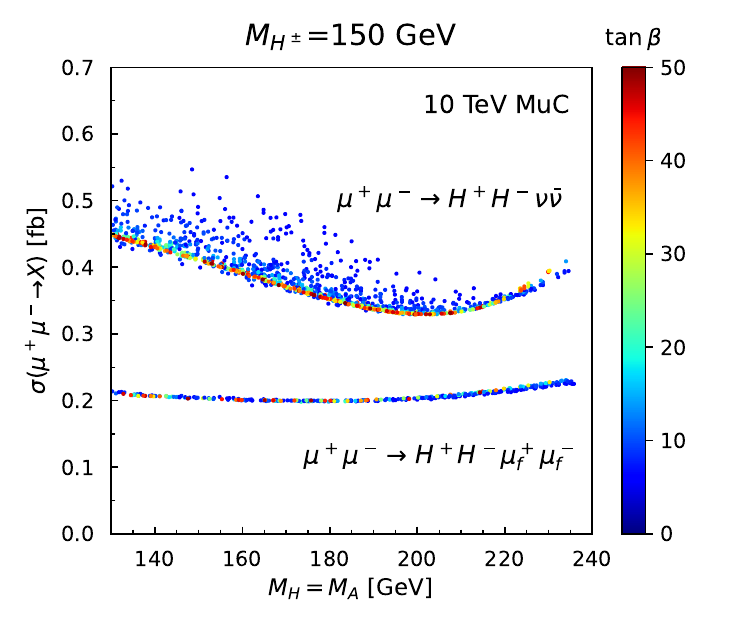}
  \caption{Parton-level cross sections as a function of $\mhha$
  for the processes  $\mmu \to H^+ H^- \nnu$ 
  and  $\mmu \to H^+ H^- \mmuf$ for $\mch=130\gev$ (upper) and $\mch=150\gev$ (lower) at c.m.~energy $\sqrt{s}=10\tev$,
  over the allowed parameter points.
  $\muf$ denotes the forward muons with $|\eta_\mu|>2.5$.
  The color scale represents the values of  $\tb$.
}
  \label{fig-xsec-nnu-mmu-MuC}
\end{figure}

Let us discuss the dependence of cross sections for $\mmu \to H^+ H^- \nnu$ and $\mmu \to H^+ H^- \mmuf$ on model parameters. Unlike the Drell-Yan process, these two processes include contributions mediated by $H$ and $A$, making their cross sections potentially sensitive to $\mhha$. Additionally, VBS contributions through $W^+ W^- \rightarrow h^* \rightarrow H^+ H^-$ to $\mmu \to H^+ H^- \nnu$ and $ZZ\to h^* \to H^+ H^-$ to $\mmu \to H^+ H^- \mmuf$ introduce dependences on $\tb$.

\autoref{fig-xsec-nnu-mmu-MuC} illustrates the parton-level cross sections 
as a function of $\mhha$ for the processes $\mmu \to H^+ H^- \nnu$ and $\mmu \to H^+ H^- \mmuf$
with $\mch = 130~\gev$ (upper panel) and $\mch = 150~\gev$ (lower panel) at $\sqrt{s} = 10~\tev$. 
The color codes denote $\tan\beta$. The analysis considers parameter points allowed by theoretical and experimental constraints discussed in Sec.~\ref{sec-review}.
The cross sections were computed using {\small\sc MadGraph5\_aMC@NLO}~\cite{Alwall:2011uj}, which includes all relevant tree-level Feynman diagrams contributing to these $2 \to 4$ processes. 

For the process involving forward muons ($\muf$), we apply the condition $2.5 < |\eta_\mu| < 4.7413$. 
The lower bound ensures that forward muons fall outside the conventional pseudorapidity coverage limit of the MuC. The upper bound, $|\eta| < 4.7413$, is introduced to regulate the divergence that occurs as the scattering angle $\theta$ approaches 0 or $\pi$, a consequence of $t$-channel photon-mediated diagrams. 
This divergence is typically canceled by higher-order QED corrections, particularly from soft photon emissions. In practice, such effects are handled by imposing a minimum scattering angle cut during data analysis. Considering the high collision energy at the MuC, we apply the angular constraint $1^\circ < \theta < 179^\circ$, which corresponds to $|\eta| < 4.7413$.

\autoref{fig-xsec-nnu-mmu-MuC} shows that the results for $\mch=130\gev$ exhibit similar trends to those for $\mch=150\gev$, with two notable differences. First, the cross sections for $\mch=130\gev$ are moderately larger than those for $\mch=150\gev$, increasing by approximately 20\%. This enhancement is due to the  lighter $\mch$. Second, the upper bounds on $\mhha$ are lower for $\mch=130\gev$ than for $\mch=150\gev$, as dictated by the theoretical constraints of perturbativity and unitarity, which prohibit excessively large mass differences between $\mch$ and $\mhha$ (see \autoref{eq:quartic}).

Another important observation from \autoref{fig-xsec-nnu-mmu-MuC} is that the cross sections for both $\mmu \to H^+ H^- \nnu$ and $\mmu \to H^+ H^- \mmuf$ exhibit only a mild dependence on the model parameters $\mhha$ and $\tb$.
Notably, the cross section for $\mmu \to H^+ H^- \mmuf$ remains nearly constant 
across the allowed parameter space, varying by only about 10\%. 
In contrast, the cross section for $\mmu \to H^+ H^- \nnu$ shows a more pronounced dependence 
on the model parameters, with variations of approximately 50\%. 
For a given $\tb$, the cross section initially decreases as $\mhha$ increases, 
reaches a minimum at $\mhha\simeq 182~ (200)\gev$ for $\mch=130 ~(150)\gev$, and then rises again.
A detailed discussion of the dominant Feynman diagrams and the dependence of $\sigma(\mmu \to H^+ H^- \nnu)$ on model parameters can be found in \autoref{appendix}.

Given these observations, we adopt a conservative approach by selecting $\mhh=\ma=200\gev$, $\tb=10$, and $m_{12}^2 = 3.84 \times 10^3 \gev^2$ for our signal-to-background analysis. This parameter point represents a pessimistic scenario, yielding the lowest cross section for $\mmu \to H^+ H^- \nu\bar{\nu}$ within the allowed parameter space with $\mch=150\gev$. If we can achieve a discovery-level signal significance for this conservative benchmark, parameter points with higher cross sections would provide better detection prospects.

Nevertheless, we note that the discovery reach could be moderately affected by different choices of model parameters through small variations in kinematics and event selection efficiencies. Such effects are more relevant at the 10 TeV MuC, where the cross section of $\mmu \to H^+ H^- \nnu$ is comparable to that of $\mmu \to H^+ H^-$. In contrast, at the 3 TeV MuC, the cross section of $\mmu \to H^+ H^-$ is an order of magnitude larger than those of $\mmu \to H^+ H^- \nnu$ and $\mmu \to H^+ H^- \mmuf$ (these cross sections will be presented in \autoref{tab-xsec-muc}). As a result, the discovery potential at 3 TeV exhibits much weaker dependence on the model parameters, since the $\mmu \to H^+ H^-$ process is independent of these parameters for a given $\mch$.

\begin{table*}[t]
  \centering
  \footnotesize
  \setlength{\tabcolsep}{3pt}
  \renewcommand{\arraystretch}{1.3}
  \begin{tabular}{|c||c|c|c||c|c|c|}
  \hline
  \multicolumn{7}{|c|}{Cross sections of the pair production of  charged Higgs bosons at the MuC}\\ \hline
   &  \multicolumn{3}{c||}{$\sqrt{s}=3\tev$} &  \multicolumn{3}{c|}{$\sqrt{s}=10\tev$}  \\ \hline
$\mch$ & $130\gev$ & $150\gev$ & $170\gev$ & $130\gev$ & $150\gev$ & $170\gev$ \\ \hline\hline
$\sg(\mmu\rightarrow H^+ H^-)$ & $3.26\fb$& $ 3.24\fb$ & $ 3.22 \fb$ & $0.294\fb$ & $0.296 \fb$ & $0.295\fb$ \\  \hline
$\sg(\mmu\rightarrow H^+ H^- \nu\bar{\nu})$ & $0.195\fb$ & $0.149\fb$ & $0.129\fb$  & $ 0.448\fb$& $0.347\fb$ & $0.303\fb$ \\  \hline
~~$\sg(\mmu\rightarrow H^+ H^- \mmuf)$~~ & $0.261\fb$ & $0.176\fb$ & $0.127\fb$ & $0.262 \fb$ & $0.204\fb$ & $0.163\fb$ 
  \\  \hline
  \end{tabular}
  \caption{Cross sections of the $\ch$ pair production for the signal at the 3 TeV and 10 TeV MuC. We set $\mhh=\ma=200\gev$, $\tb=10$, and $m_{12}^2 = 3.84 \times 10^3 \gev^2$. For the $\ch$ pair production associated with $\nu\bar{\nu}$, all three neutrino flavors are included. $\muf$ denotes the forward muons with $|\eta_\mu|>2.5$.}
  \label{tab-xsec-muc}
\end{table*} 

We now compare the cross sections of the three production channels 
in \autoref{eq-signal-DY-muc}, \autoref{eq-signal-CC-muc}, and  \autoref{eq-signal-NC-muc}
for the charged Higgs mass \begin{equation}
\label{eq-mch}
\mch=130, \;150, \;170 \gev.
\end{equation}
\autoref{tab-xsec-muc} presents the parton-level cross sections at 3 TeV and 10 TeV MuC
for $\mhha=200\gev$, $\tb=10$, and $m_{12}^2 = 3.84 \times 10^3 \gev^2$.
We used \textsc{MadGraph5\_aMC@NLO} version 3.5.0.

The Drell-Yan process shows nearly constant cross sections for different $\mch$ values at a given $\sqrt{s}$: approximately 3 fb at $\sqrt{s}=3\tev$ and 0.3 fb at $\sqrt{s}=10\tev$. This stability arises because the chosen $M_{H^\pm}$ values (see \autoref{eq-mch}) are relatively small compared to the multi-TeV collision energies.  The significant decrease in cross sections from $\sqrt{s}=3\tev$ to $\sqrt{s}=10\tev$ reflects the typical behavior of Drell-Yan processes, where cross sections scale as $1/s$. In contrast, the cross sections for $\mmu\to H^+ H^-\nu \bar{\nu}$ and $\mmu\to H^+ H^- \muf^+\muf^-$ moderately increase with $\sqrt{s}$, following the characteristic $\log^2({s}/{m_V^2})$ dependence of the cross sections for VBS processes~\cite{Han:2021udl}.

The relative importance of the three production channels varies with collision energy. 
At $\sqrt{s}=3\tev$, the Drell-Yan process dominates, yielding the highest cross section. 
Meanwhile, the processes involving two forward muons and those with two neutrinos exhibit comparable cross sections.
However, this hierarchy changes at $\sqrt{s}=10\tev$. 
In this higher energy regime, $\mmu \rightarrow H^+ H^- \nnu$ emerges as the dominant process, 
exhibiting the largest cross section. 
The Drell-Yan process $\mmu \rightarrow H^+ H^-$
becomes the second most significant,
while $\mmu \rightarrow H^+ H^- \nu\bar{\nu}$ now shows the smallest cross section among the three channels.
Despite this, we will demonstrate that even at $\sqrt{s}=10\tev$,
the Drell-Yan production remains dominant after the final selection.

Next, we present the signal-to-background analysis at the detector level.\footnote{In our analysis, we did not include initial state radiation (ISR) or beamstrahlung effects, as these are significantly less pronounced at a muon collider compared to an $e^+e^-$ collider. The much larger muon mass suppresses both ISR and beamstrahlung, making such corrections less critical for our study.}
We conducted showering with \textsc{Pythia} version 8.307. 
A rapid detector simulation was performed using \textsc{Delphes} version 3.5.0, 
utilizing the \texttt{delphes\_card\_\\MuonColliderDet.tcl} card.
 The MuC \textsc{Delphes} card accommodates slightly different values 
for the $\tau$ tagging and mistagging rates compared to those for the HL-LHC: 
for $p_T \geq 10\gev$, $P_{\tau\to\tau} \simeq 80\%$, $P_{e\rightarrow \tau} \simeq 0.1\%$, 
and $P_{j\to\tau}\simeq 2\%$.

For jet clustering, we employed the exclusive Valencia algorithm~\cite{Boronat:2014hva,Boronat:2016tgd}, as implemented in \texttt{FastJet}. The Valencia algorithm is particularly effective at high-energy lepton colliders because it efficiently handles initial state radiation and beam-induced backgrounds (BIBs) by explicitly incorporating beam jets.

We specifically chose the \emph{exclusive} clustering mode because our signal ($H^+H^- \rightarrow t^* b \tau\nu$) produces a fixed number of jets—exactly five. The exclusive mode halts clustering once the predetermined jet multiplicity is reached, ensuring optimal performance for signals with a known jet count. In contrast, the inclusive clustering algorithm tends to misidentify or merge closely spaced signal jets, especially at multi-TeV muon colliders, where jets from the same parent particle ($H^\pm$) become highly collimated.

The exclusive Valencia algorithm has three configurable parameters: the jet radius \( R \), and the energy and angular weighting parameters \( \beta \) and \( \gamma \) that enter the clustering distance measures~\cite{Boronat:2016tgd}. Specifically, \( \beta \) and \( \gamma \) define the inter-particle distance \( d_{ij} \) and the beam distance \( d_{iB} \) as:
\begin{align} 
d_{ij} &= \frac{2\,\text{min}(E_i^{2\beta}, E_j^{2\beta}) (1 - \cos\theta_{ij})}{R^2}, \\
d_{iB} &= E_i^{2\beta} \sin^{2\gamma} \theta_{iB},
\end{align}
where \( \theta_{iB} \) is the angle of particle \( i \) relative to the beam axis.  
For robust jet reconstruction, we adopt the canonical settings for the algorithm: \( R = 0.2 \) and \( \beta = \gamma = 1 \).

To identify relevant SM backgrounds for the $bbjj\tau\nu$ final state, we note that the $\tau\nu$ system must originate from a $W^\pm$ boson at the MuC. Since $W^\pm$ bosons are produced in pairs at the MuC, the two light jets in the final state must arise from the decay of the second $W^\pm$. This makes $\mu^+ \mu^- \rightarrow W^+ W^- b\bar{b}$ the primary background. Notably, this process also includes top quark pair production as a subset.

There are two additional backgrounds worth considering. The first is $\mu^+ \mu^- \rightarrow W^+ W^- jj$, where light quark jets are misidentified as $b$ jets. This includes contributions from $\mu^+ \mu^- \rightarrow W^+ W^- Z$ with $Z \to jj$. The second is $\mu^+ \mu^- \rightarrow t\bar{t}Z$, followed by $Z \to \nu\bar{\nu}$.

However, both of these additional backgrounds are negligible compared to the dominant $W^+ W^- b\bar{b}$ process. For the $W^+ W^- jj$ background, the contribution is strongly suppressed by realistic $b$-jet mistagging rates. Specifically, the cross section for $\mu^+ \mu^- \rightarrow W^+ W^- jj \rightarrow \tau\nu jj bb$, with kinematic cuts of $p_T > 20~\gev$, $|\eta| < 2.5$, and $\Delta R_{ij} > 0.4$, is approximately $1.9\ab$ at the 3 TeV MuC and $0.4\ab$ at the 10 TeV MuC. 
For the $t\bar{t}Z \to bbjj\tau\nu\nu\bar{\nu}$ background, the corresponding cross sections under the same kinematic cuts are approximately $5.3\ab$ at 3 TeV and $0.027\ab$ at 10 TeV. In contrast, the cross sections for $\mu^+ \mu^- \to W^+ W^- b\bar{b}$ after the same selection are $549\ab$ at 3 TeV and $68.9\ab$ at 10 TeV. 
Given this large disparity, we identify $\mu^+ \mu^- \rightarrow W^+ W^- b\bar{b}$ as the dominant background for the $bbjj\tau\nu$ final state at the MuC.

Brief comments on BIBs are in order here.
Beam-induced backgrounds arise from the decay of muons in the beam, which produces electrons and positrons. As these interact with machine components, various secondary particles such as photons, electron-positron pairs, hadrons, and neutrinos are generated. Most of these secondary particles result from forward scattering, causing them to be directed along the beam and to have low energies, typically below 1 GeV~\cite{Bartosik:2024ulr}. 
Given that our final state requires five hard jets in the central region, 
BIBs are of negligible concern. Thus, we do not consider BIBs a significant background source for our analysis of the $bbjj\tau\nu$ final state at the MuC.

We establish the following basic selection criteria:
\begin{itemize}
  \item Jet multiplicity: Exactly 2 light jets and 2 $b$-jets, 
  along with one hadronically decaying tau lepton, 
  with transverse momentum $p_T > 25$ GeV and pseudorapidity $|\eta| < 2.5$.
  \item Missing transverse energy: $E_T^{\rm miss} > 50$ GeV.
  \item Lepton veto: Events containing any electrons or muons with $p_T > 10$ GeV and $|\eta| < 2.5$ are vetoed to suppress backgrounds from leptonic decays of $W$ bosons.
\end{itemize}

\begin{figure*}[!t]
  \centering
  \includegraphics[width=\textwidth]{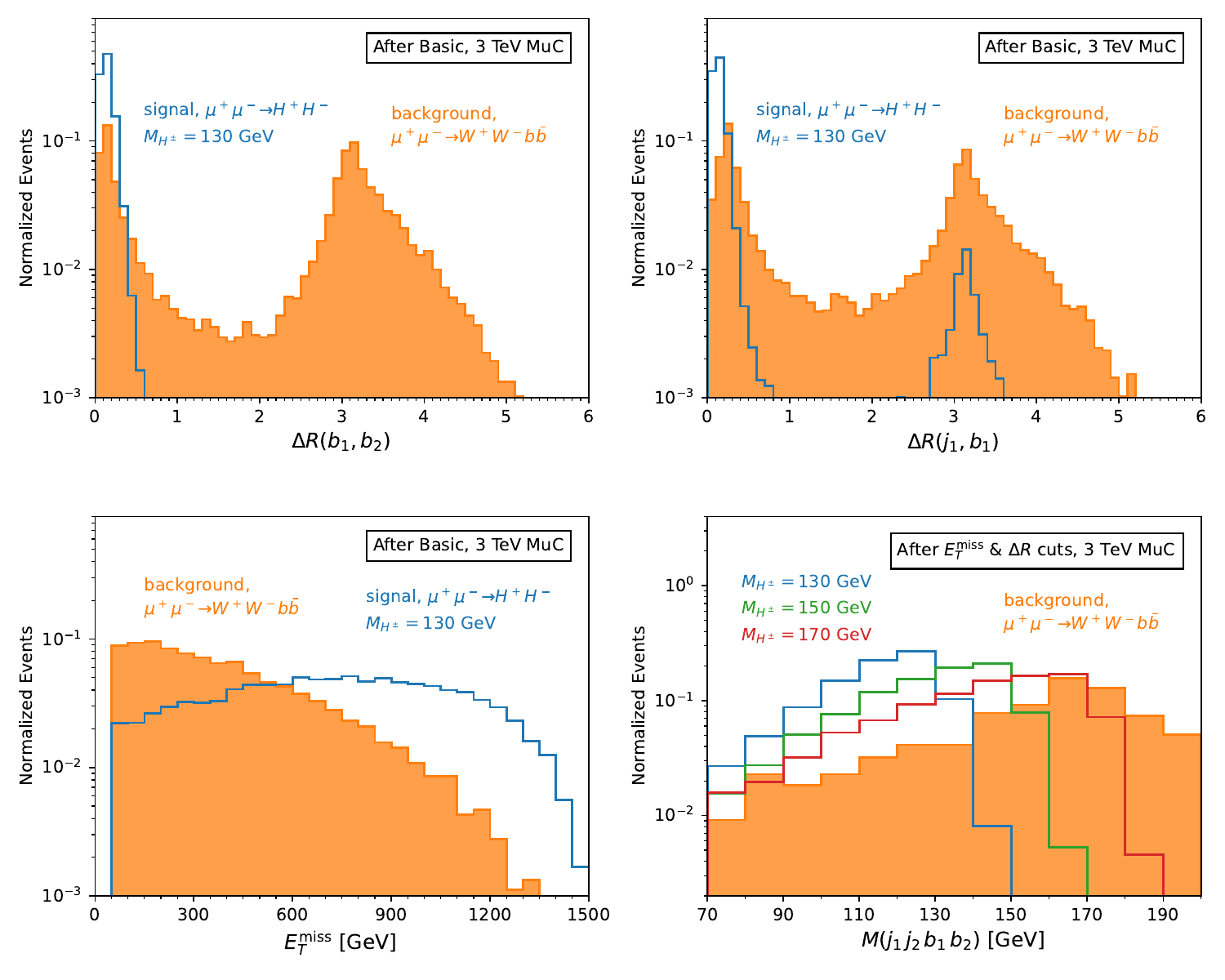}
  \vspace{-0.7cm}
  \caption{Normalized distributions for $\Delta R(b_1, b_2)$ (upper-left), $\Delta R(j_1, b_1)$ (upper-right), 
$\met$ (lower-left),
and $M(j_1 j_2 b_1 b_2)$ (lower-right) for the signal (solid lines) and background (orange histograms).
For  the $\Delta R$ and $\met$ distributions,
we show the results for $\mch=130\gev$ after applying the basic selection criteria,
while for $M(j_1 j_2 b_1 b_2)$, we show the results for $\mch=130\gev$ (blue), 150 GeV (green), and 170 GeV (red) after additionally imposing $\met>500\gev$, $\Delta R(b_1,b_2) < 0.6$
and $\Delta R(j_i,b_{i'})|_{i,i'=1,2}<0.6$.
}
  \label{fig-kin-muc}
\end{figure*}

To establish an effective cut-flow, we begin with a detailed analysis of the key kinematic variables that differentiate signal from background. 
The normalized distributions of the critical variables are presented in \autoref{fig-kin-muc}:
$\Delta R(b_1, b_2)$ in the upper-left panel, 
$\Delta R(j_1, b_1)$ in the upper-right panel, $\met$ in the lower-left panel, 
and $M(j_1 j_2 b_1 b_2)$ in the lower-right panel. 
The signal distributions are depicted as solid lines, 
while the background distributions are shown as orange histograms. 
The $\Delta R$ and $\met$ distributions are presented after applying the basic selection criteria, for the representative case of $\mch = 130$ GeV.\footnote{The $\Delta R$ and $\met$ distributions remain almost identical for $\mch=130,~150,~170\gev$ since these masses are much smaller than $\sqrt{s}$ at the multi-TeV MuC.}
For the $M(j_1 j_2 b_1 b_2)$ distribution,
we show the results for $\mch=130\gev$ (blue), 150 GeV (green), and 170 GeV (red)
after imposing 
\begin{align}
&\Delta R(b_1, b_2)<0.6, \; \Delta R(j_i,b_{i'})\big|_{i,i'=1,2}<0.6,
\\ \nn & \met>500\gev.
\end{align}

The $\Delta R(b_1, b_2)$ distribution in \autoref{fig-kin-muc} clearly shows that 
this variable is highly discriminating for separating the signal from the background. 
In the signal events, the two $b$-jets are typically adjacent, 
originating from the decay of the same parent $H^\pm$. 
In contrast, the background exhibits a broader distribution 
with larger angular separations between the two $b$-jets. 
A stringent criterion of $\Delta R(b_1, b_2) < 0.6$ is effective in isolating the signal from the background.

Similarly, the angular separation between the leading light jet and the leading $b$-jet, $\Delta R(j_1, b_1)$, exhibits a distinct pattern for the signal compared to the background. The signal distribution features a prominent primary peak at $\Delta R(j_1, b_1) \simeq 0.1$, indicating that $j_1$ and $b_1$ originate from the same $H^\pm$ decay. This trend is consistently observed in other combinations, such as $\Delta R(j_1, b_2)$, $\Delta R(j_2, b_1)$, and $\Delta R(j_2, b_2)$, as expected from the kinematics of the signal process. 

Interestingly, the signal distribution also exhibits a secondary peak at larger values around $\Delta R \sim 3$. 
We find that this arises from occasional misidentification between light jets and tau-jets, which is accounted for in our detector simulation. Specifically, the {\sc\small Delphes} setup incorporates realistic mistagging probabilities, including a $P_{j\to\tau} \simeq 2\%$ probability for a light jet to be misidentified as a tau jet, leading to this feature in the distribution.

Another crucial discriminating variable is the missing transverse energy $\met$, shown in the lower-left panel of \autoref{fig-kin-muc}. The signal exhibits a significantly higher $\met$ distribution compared to the background, attributed to the different production mechanisms and mother particles of the neutrino. In the signal process, the $[\tau\nu]$ system originates from the decay of a charged Higgs boson
produced through a $2 \rightarrow 2$ scattering process at the MuC. The large energy transfer to the $H^\pm$ results in a more energetic neutrino from its subsequent decay, leading to substantial missing transverse energy in the final state. In contrast, the background process involves a $W^\pm$ boson produced through a $2 \rightarrow 4$ scattering process, which inherently results in a softer energy transfer to the $W^\pm$ and, consequently, a less energetic neutrino from its decay. As a result, a stringent cut of $\met > 500$ GeV is highly effective in enhancing the signal significance.

Finally, we present the invariant mass distribution constructed from the two leading $b$-jets and two leading light quark jets in the lower-right panel of \autoref{fig-kin-muc} for $m_{H^\pm} = 130$ GeV (blue), 150 GeV (green), and 170 GeV (red). To demonstrate the efficiency of our proposed cuts on $\met$ and $\Delta R$ in revealing the invariant mass peaks over the background distributions, we show the distributions after imposing the requirements $\Delta R(b_1, b_2) < 0.6$, $\Delta R(j_i, b_{i'}) < 0.6$ for $i, i' = 1, 2$, and $\met > 500$ GeV. We observe distinct resonance peaks in the signal distributions, although the peak positions appear slightly below the true charged Higgs boson mass, primarily due to smearing effects in reconstructing the two $b$-jets and two light jets from the $H^\pm$ decays. The background distribution also exhibits a peak marginally above the top quark mass, largely attributable to contributions from top quark pair production.

\begin{table*}[h]
  \centering
  \footnotesize
  \setlength{\tabcolsep}{3pt}
  \renewcommand{\arraystretch}{1.3}
  \begin{tabular}{|c|c|c|c|c|c|c|c|c|c|c|}
  \hline
  \multicolumn{8}{|c|}{Cut-flow for $\mu^+ \mu^- \rightarrow H^+ H^- \rightarrow t^* b \tau \nu \rightarrow j j b b \tau \nu$ at a 3 TeV MuC with $\mathcal{L}_\text{tot}=1\iab$ }\\ \hline
 \multirow{2}{*}{Cut} & \multirow{2}{*}{$\sigma_{\rm bg}$ [fb]} & \multicolumn{2}{c|}{$\mch=130\gev$} & \multicolumn{2}{c|}{$\mch=150\gev$} & \multicolumn{2}{c|}{$\mch=170\gev$}  \\ \cline{3-8}
  & & $\sigma_{\rm sg}$ [fb] & $\mathcal{S}^{10\%}_{1 {\rm ab}^{-1}}$ & $\sigma_{\rm sg}$ [fb] & $\mathcal{S}^{10\%}_{1 {\rm ab}^{-1}}$ & $\sigma_{\rm sg}$ [fb] & $\mathcal{S}^{10\%}_{1 {\rm ab}^{-1}}$ \\ \hline
  %Basic
  Basic & ~$5.49\times 10^{-1}$~  
  	& ~$1.04 \times 10^{-1}$~ & $1.65$ 
	& ~$1.04 \times 10^{-1}$~  & $1.65$ 
	& ~$3.66 \times 10^{-2}$~ & 0.600\\ \hline
  %met
  $\met>500\gev$  & $1.67\times 10^{-1}$  
  	& $7.49 \times 10^{-2}$ & $3.19$ 
	& $7.36 \times 10^{-2}$  & $3.14$  
	& $2.61 \times 10^{-2}$ & 1.19 \\ \hline
  %DRb1b1
  $\Delta R (b_1,b_2)<0.6$  & $5.75\times 10^{-2}$ 
  	& $7.48 \times 10^{-2}$ & $6.33$ 
	& $7.34 \times 10^{-2}$ & 6.23 
	& $2.60 \times 10^{-2}$ & 2.50 \\ \hline
   %DRj_1,b_1
  $\Delta R (j_1,b_1)<0.6$  & $2.97 \times 10^{-2}$ 
  	& $7.19 \times 10^{-2}$ & $8.48$ 
	& $7.08 \times 10^{-2}$ & 8.38 
	& $2.50 \times 10^{-2}$ & 3.50 \\ \hline
   %DRj_1,b_2
    $\Delta R (j_1,b_2)<0.6$  & $2.87 \times 10^{-2}$ 
    	& $7.18 \times 10^{-2}$ & $8.62$ 
	& $7.06 \times 10^{-2}$ & 8.50 
	& $2.50 \times 10^{-2}$ & 3.57  \\ \hline
   %DRj_2,b_1
  $\Delta R (j_2,b_1)<0.6$  & $1.29 \times 10^{-2}$ 
  	& $6.82 \times 10^{-2}$ & $11.3$ 
	& $6.73 \times 10^{-2}$  &  11.2 
	& $2.38 \times 10^{-2}$ & 4.95  \\ \hline
     %DRj_2,b_2
$\Delta R (j_2,b_2)<0.6$  & $1.22 \times 10^{-2}$ 
	& $6.79 \times 10^{-2}$ & $11.4$ 
	& $6.66 \times 10^{-2}$  & 11.3 
	& $2.34 \times 10^{-2}$ & 4.99 \\ \hline
     %M
$M(j_1j_2b_1b_2)<m_t$  & $6.61 \times 10^{-3}$ 
	& $6.74 \times 10^{-2}$ &$13.7$  
	& $6.61 \times 10^{-2}$ & 13.5 
	& $2.24 \times 10^{-2}$ & 6.06 \\ \hline
  \end{tabular}
  \caption{Cut-flow for the signal process $\mu^+ \mu^- \rightarrow H^+ H^- \rightarrow t^* b \tau \nu \rightarrow j j b b \tau \nu$ with $M_{H^\pm} = 130$, 150, and 170 GeV at the 3 TeV MuC. We set $\mhh = \ma = 200$ GeV, $\tan\beta = 10$, and $m_{12}^2 = 3.84 \times 10^3$ GeV$^2$. The dominant background process is $\mu^+ \mu^- \to W^+ W^- b \bar{b}$. The significance $\mathcal{S}^{10\%}_{1 {\rm ab}^{-1}}$ is calculated assuming a 10\% background uncertainty and an integrated luminosity of 1 ab$^{-1}$.
}
  \label{tab:cutflow:3TeV:muc}
\end{table*}

Based on the observations from the key kinematic variable distributions, we propose an effective cut-flow strategy to achieve the high discovery potential at the 3 TeV MuC, 
as summarized in \autoref{tab:cutflow:3TeV:muc}. 
The significances are calculated for a total integrated luminosity of $1\iab$ 
and a 10\% background uncertainty. 
Although we do not present the cut-flow for the negligible $\mmu\to H^+H^- \nnu$ and $\mmu\to H^+H^- \mmuf$, their contributions are included in calculating the significance.
In the following discussion, the selection efficiency of each specific cut is measured relative to the preceding cut in the cut-flow.

After applying the basic selection criteria, the signal cross sections show similar values across the three $\mch$ cases of 130, 150, and 170 GeV, with a slight decrease as $\mch$ increases. The first effective kinematic cut is on the missing transverse energy, requiring $\met > 500$ GeV. This cut achieves approximately 70\% selection efficiency for the signal while retaining only about 30\% of the background, providing substantial initial background suppression.

The most crucial cut is on the angular distance between the leading and subleading $b$-jets, $\Delta R(b_1, b_2) < 0.6$. Among events that satisfy the $\met$ cut, almost all signal events survive this $\Delta R(b_1, b_2)$ cut, while only about 30\% of background events remain. At this stage, we achieve substantial signal significances: $\mathcal{S}^{10\%}_{1 {\rm ab}^{-1}} \approx 6$ for $m_{H^\pm} = 130$ GeV and $150$ GeV, and $\mathcal{S}^{10\%}_{1 {\rm ab}^{-1}} \approx 2.5$ for $m_{H^\pm} = 170$ GeV. The smaller significance for $m_{H^\pm} = 170$ GeV is due to the reduced production cross section at higher charged Higgs boson mass.

Subsequently, we impose a series of angular distance cuts between a $b$ jet and a light jet, 
collectively denoted as $\Delta R(j_i, b_{i'})|_{i, i' = 1, 2} < 0.6$. 
Each of these cuts consistently reduces the background while retaining almost all of the signal events. The selection efficiency of this series of cuts is about 20\% for the background and about 90\% for the signal.

Finally, we impose a condition on the invariant mass of the two $b$-jets and two light jets, such that $M(b_1 b_2 j_1 j_2) < 173$ GeV. We apply an upper bound on $M(b_1 b_2 j_1 j_2)$ rather than a mass window. This approach is designed to retain as many signal events as possible, since the signal cross sections after the final selection are only of the order of $\mathcal{O}(10)$ ab, resulting in dozens of signal events with a total integrated luminosity of $1$ ab$^{-1}$. Additionally, the relatively broad  resonance peak, due to smearing effects in reconstructing the two $b$-jets and two light jets, is another reason for not imposing a narrow mass window (see \autoref{fig-kin-muc}). This final selection ensures a significant discovery potential for all three benchmark cases. For $M_{H^\pm} = 130$, 150, and 170 GeV, the final signal significances are 13.7, 13.5, and 6.06, respectively.

To explore how higher collision energy affects the sensitivity to the $\ch\rightarrow t^* b$ mode, we analyze the 10 TeV MuC with a total integrated luminosity of $10\iab$. One might expect enhanced signal sensitivity compared to the 3 TeV MuC, given both the larger luminosity and the increased importance of VBS processes ($\mu^+\mu^- \rightarrow H^+ H^-\nnu/\mmuf$): see \autoref{tab-xsec-muc}.

\begin{table*}[h]
  \centering
  \footnotesize
  \setlength{\tabcolsep}{3pt}
  \renewcommand{\arraystretch}{1.3}
  \begin{tabular}{|c|c|c|c|c|c|c|c|c|c|c|}
  \hline
  \multicolumn{8}{|c|}{Cut-flow for $\mu^+ \mu^- \rightarrow H^+ H^- \rightarrow t^* b \tau \nu \rightarrow j j b b \tau \nu$ at a 10 TeV MuC with $\mathcal{L}_\text{tot}=10\iab$ }\\ \hline
 \multirow{2}{*}{Cut} & \multirow{2}{*}{$\sigma_{\rm bg}$ [fb]} & \multicolumn{4}{c|}{$\mch=130\gev$} &  \multicolumn{2}{c|}{$\mch=170\gev$}  \\ \cline{3-8}
  & & $\sigma_{\rm sg}^{\rm DY}$ [fb] & $\sigma_{\rm sg}^{\nnu}$ [fb] & $\sigma_{\rm sg}^{\muf\muf}$ [fb] & $\mathcal{S}^{10\%}_{10 {\rm ab}^{-1}}$ & $\sigma_{\rm sg}^{\rm DY}$ [fb] & $\mathcal{S}^{10\%}_{10 {\rm ab}^{-1}}$ \\ \hline
  %Basic
 Basic &  \;$6.89\times 10^{-2}$\;
 	& $9.49\times 10^{-3}$ & $ 5.28\times 10^{-3}$  & $2.39 \times 10^{-3}$  &  2.17
		& $3.14 \times 10^{-3}$ & 0.677 \\ \hline
%MET
  $\met>1\tev$  &  $3.10\times 10^{-2}$
  	& $7.85 \times 10^{-3}$ & $6.87 \times 10^{-4}$ & $5.25 \times 10^{-5}$ &  2.24
		& $2.62 \times 10^{-3}$ & 0.773 \\ \hline
%DR bb
  $\Delta R (b_1,b_2)<0.2$  &  $1.58 \times 10^{-2}$
  	& $7.71 \times 10^{-3}$ & $ 5.53\times 10^{-4}$  & $4.64 \times 10^{-5}$ & 3.65 
		& $2.54 \times 10^{-3}$ & 1.28 \\ \hline 
%DR jb
  $\Delta R (j_i,b_{i'})<0.2$  &  $8.95 \times 10^{-4}$
  	& $6.25 \times 10^{-3}$ & $2.11 \times 10^{-4}$ & $2.38 \times 10^{-5}$ &12.2  
		& $2.06 \times 10^{-3}$ & 5.17 \\ \hline
%M
  $M(j_1j_2b_1b_2)< m_t$  &  $4.21\times 10^{-4}$
  	& $6.19 \times 10^{-3}$ & $2.10\times 10^{-4}$ & $2.38 \times 10^{-5}$ & 14.8  
		& $1.89 \times 10^{-3}$ & 6.26  \\ \hline
  \end{tabular}
  \caption{
  Cut-flow for the signal process $\mu^+ \mu^- \rightarrow H^+ H^- \rightarrow t^* b \tau \nu \rightarrow j j b b \tau \nu$ at the 10 TeV MuC with a total integrated luminosity of 10 ab$^{-1}$. The dominant background process is $\mu^+ \mu^- \to W^+ W^- b \bar{b}$.  For $M_{H^\pm} = 130$ GeV, we present the detailed cut-flow of the cross section for the Drell-Yan process ($\sigma_{\rm sg}^{\rm DY}$), $\mmu\rightarrow H^+H^-\nnu$ ($\sigma_{\rm sg}^{\nnu}$), and $\mmu\rightarrow H^+H^-\mmuf$ ($\sigma_{\rm sg}^{\muf\muf}$). For $\mch = 170$ GeV, we present the cut-flow of the dominant $\sigma_{\rm sg}^{\rm DY}$. The cut $\Delta R (j_i,b_{i'})<0.2$ collectively denotes four possible combinations with $i,i'=1,2$.
 }
  \label{tab:cutflow:10TeV:muc}
\end{table*}

In \autoref{tab:cutflow:10TeV:muc}, 
we present the cut-flow of the cross sections for $\mch=130\gev$ and $\mch=170\gev$ at the 10 TeV MuC. 
The results for $\mch=150\gev$ are omitted as they are very similar to those for $\mch=130\gev$, 
with differences below 1\%. 
The significances are calculated for a total integrated luminosity of $10\iab$ 
and a 10\% background uncertainty. 
For $M_{H^\pm}=130\gev$, we present the cut-flow for all three production channels: 
the Drell-Yan process, $\mmu\rightarrow H^+H^-\nnu$, and $\mmu\rightarrow H^+H^-\mmuf$, 
with their respective cross sections denoted by $\sigma_{\rm sg}^{\rm DY}$, $\sigma_{\rm sg}^{\nnu}$, 
and $\sigma_{\rm sg}^{\muf\muf}$. 
After the basic selection, the cross sections for the three processes are of the same order of magnitude.  
However, the $\met > 1\tev$ cut substantially suppresses 
$\sigma_{\rm sg}^{\nnu}$ and $\sigma_{\rm sg}^{\muf\muf}$, leaving less than 10\% of events. 
In contrast, the Drell-Yan process maintains a high selection efficiency, around 83\%. 
In the case of $\mch=170\text{ GeV}$,
we present only the Drell-Yan cross sections
since the other processes are negligible after the final selection. 
However, the signal significances are calculated based on the total cross section from all three processes.

The most decisive cuts for suppressing the background are the series of $\Delta R (j_i, b_{i'}) < 0.2$, which results in a background selection efficiency of about 5.7\%. These cuts achieve a significance well above the $5\sigma$ discovery threshold. Our final selection on the invariant mass of two $b$ jets and two light jets further enhances the significance. For $\mch=130\gev$, $150\gev$, and $170\gev$, the significances are 14.8, 14.4, and 6.26, respectively.

Despite these results, we conclude that the 3 TeV MuC is more efficient than the 10 TeV MuC for probing the new physics signal of $\ch\rightarrow t^* b$. 
Although both the 3 TeV and 10 TeV MuCs achieve similar signal significances, the 3 TeV MuC requires only $1\iab$ luminosity, whereas the 10 TeV MuC demands a much higher luminosity of $10\iab$.

A discussion of CLIC's discovery potential for the $\ch\to t^* b$ mode merits attention. In its final stage, CLIC~\cite{Brunner:2022usy} will operate at a c.m.~energy of 3 TeV with an integrated luminosity of approximately $5\iab$. Unlike a multi-TeV MuC, where detection is limited to $|\eta| < 2.5$ due to BIBs, CLIC faces no such restrictions as an electron-positron collider. As a result, a 3 TeV CLIC is expected to have superior sensitivity to the $H^\pm \to t^* b$ mode compared to a 3 TeV MuC. The intermediate stage at $\sqrt{s} = 1.5\tev$ should also provide substantial discovery reach. Given the distinct collider characteristics, a dedicated study of this channel at CLIC would be valuable for future study.

\section{Conclusions}
\label{sec-conclusions}

In this study, we presented the first comprehensive analysis investigating the discovery potential of future colliders for the decay channel of the light charged Higgs boson into an off-shell top quark and a bottom quark, $H^\pm \rightarrow t^* b$, within the context of the Type-I two-Higgs-doublet model.
 We focused on the mass range of 130 to 170 GeV, where the $H^\pm \rightarrow t^* b$ decay becomes particularly significant. To probe this new physics signal without introducing dependence on specific model parameters, we identified pair production of charged Higgs bosons as a robust production channel, followed by the decay $H^+ H^- \rightarrow t^* b \tau \nu \rightarrow bbjj\tau\nu$.

We conducted a detailed signal-to-background analysis at the HL-LHC and a prospective 100 TeV proton-proton collider, employing both comprehensive cut-flow strategies and the Boosted Decision Tree (BDT) method at the detector level. However, the results showed that the signal significance remains well below the threshold for confident detection at these colliders, primarily due to the inherent softness of the $b$ jets in the decay process, which struggle to meet basic jet clustering thresholds.

Given these constraints at hadron colliders, we extended our analysis to explore the discovery potential of a multi-TeV muon collider (MuC) for the $H^+ H^- \rightarrow t^* b \tau \nu \rightarrow bbjj\tau\nu$ signal process. The MuC offers significant advantages by fully exploiting the beam energy in fundamental particle collisions. Our analysis demonstrated that a MuC, particularly at a center-of-mass energy of 3 TeV, provides a promising environment for probing the $H^\pm \rightarrow t^* b$ decay mode. The cut-flow analysis at the 3 TeV MuC achieved a high signal significance, surpassing the $5\sigma$ discovery threshold with a total integrated luminosity of 1 ab$^{-1}$. Specifically, for $M_{H^\pm} = 130$, 150, and 170 GeV, the signal significances are 13.7, 13.5, and 6.06, respectively. In contrast, the 10 TeV MuC, despite its higher collision energy, requires a substantially larger integrated luminosity of 10 ab$^{-1}$ to achieve comparable results due to the reduced signal cross sections of the Drell-Yan production process at higher energies.

The results of our study underscore both the challenges and opportunities in searching for the light charged Higgs boson via the $t^* b$ decay mode. While high-energy hadron colliders face significant obstacles due to the soft $b$ jets, the multi-TeV MuC emerges as a highly effective platform. The detailed simulation and cut-flow strategy developed in this work provide a robust framework for future experimental searches for this promising new signal, emphasizing the critical role of a multi-TeV MuC in advancing beyond the Standard Model physics.

\subsection*{Acknowledgements}

We thank Kingman Cheung and Chih-Ting Lu for useful discussions.
And the work of D.W., J.K., P.S., and J.S. is supported by
the National Research Foundation of Korea, Grant No.~NRF-2022R1A2C1007583.
The work of S.L. is supported by Basic Science Research Program through the National Research Foundation of Korea(NRF) funded by the Ministry of Education(RS-2023-00274098).

\appendix

\section{Feynman Diagram Analysis of $\mmu \to H^+ H^- \nnu$}
\label{appendix}

\begin{figure}
\centering
\includegraphics[width=0.5\textwidth]{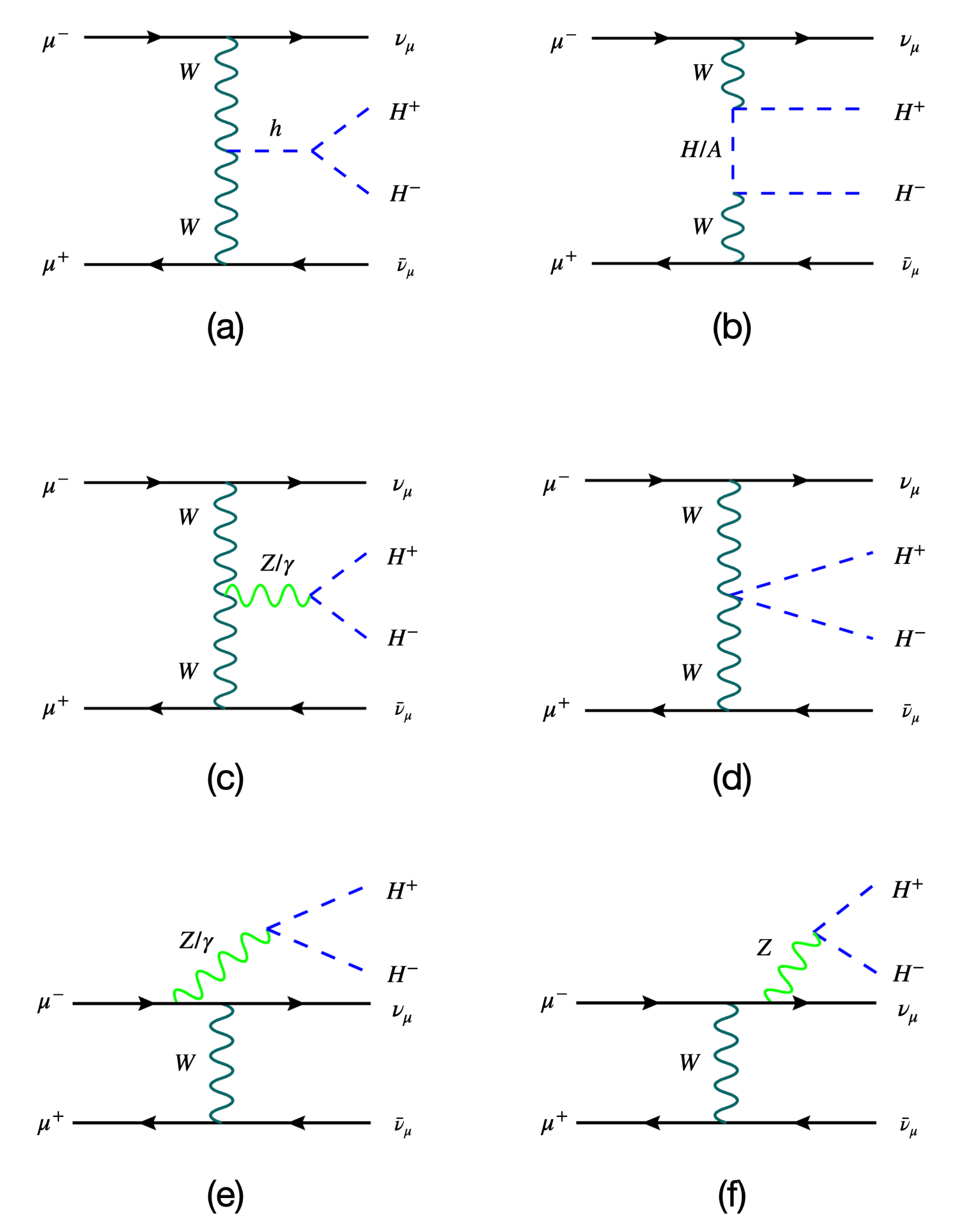}
\caption{Representative Feynman diagrams for the process $\mmu\to H^+ H^-\nnu$,
which significantly contribute to the total cross section.}
\label{fig-Feynman-mumu-H+H-nunu}
\end{figure}

In this appendix, we discuss the process:
\beq
\mu^+ \mu^- \to H^+ H^- \nu\bar{\nu},
\eeq
where $\nu$ includes all three neutrino flavors. As a $2 \to 4$ scattering process, this reaction involves 61 distinct Feynman diagrams in the Higgs alignment limit of the Type-I 2HDM. In counting these diagrams, we have neglected the Yukawa couplings involving $H^\pm$-$e^\pm$-$\nu_e$ and $H^\pm$-$\mu^\pm$-$\nu_\mu$, given their negligible contributions. This large number of diagrams complicates the analysis of how the total cross section depends on the model parameters.

To efficiently compute the contributions of individual Feynman diagrams to the total cross section, as well as their interference effects, we used {\small\sc CalcHEP}~\cite{Belyaev:2012qa}, which is well-suited for this type of analysis.
 We found that six classes of diagrams provide significant contributions, while the remaining diagrams have negligible effects. The dominant Feynman diagrams are illustrated in \autoref{fig-Feynman-mumu-H+H-nunu}:
(a) Vector boson scattering (VBS) via $W^+ W^- \rightarrow h^* \rightarrow H^+ H^-$;
(b) VBS via $t$-channel exchange of $H$ and $A$;
(c) VBS via $W^+ W^- \rightarrow Z^/\gamma^ \rightarrow H^+ H^-$;
(d) Direct VBS contribution from $W^+ W^- \rightarrow H^+ H^-$;
(e) $Z^/\gamma^ \to H^+ H^-$ radiation from the initial $\mu^-$ or $\mu^+$;
(f) $Z^* \to H^+ H^-$ radiation from the final-state $\nu_\mu$ or $\bar{\nu}_\mu$.

Among these, diagram (b) exhibits significant sensitivity to $\mhha$. However, this diagram strongly interferes destructively with diagrams (c) and (d) across the allowed parameter space, resulting in only a weak dependence of the total cross section on $\mhha$. Additionally, diagrams (e) and (f) also interfere destructively with diagrams (b), (c), and (d). For our benchmark scenario with $\mhh=\ma=200\gev$, $\tb=10$, and $m_{12}^2 = 3.84 \times 10^3 \gev^2$, diagram (a) dominates due to minimal interference effects. Nevertheless, for other parameter choices, interference significantly alters the relative contributions of these diagrams, making it challenging to identify a universally dominant Feynman diagram.

\end{document}